\documentclass[aps,prd,preprint,tightenlines,nofootinbib,byrevtex]{revtex4}
\usepackage{amssymb}
\usepackage{mathrsfs}
\usepackage{slashed}
\usepackage{graphicx,color}
\usepackage{amsmath}
\usepackage{subeqnarray}
\usepackage{multirow}
\allowdisplaybreaks

\begin{document}
\title{Majorana neutrino mass matrices with three texture zeros and the sterile neutrino}
\author{Yongchao Zhang}
\email{yczhang@pku.edu.cn}
\affiliation{Center for High Energy Physics, Peking University, Beijing, 100871, P. R. China}
\date{\today}
\begin{abstract}
  As a consequence of the LSND anomaly and other hints of an eV scale sterile neutrino from particle physics and cosmology, the neutrino sector of the standard model of particle physics has to be extended and the smallest extension is the (3+1) model, i.e., three active neutrinos plus one sterile one. In this work we study the neutrino mass matrix $M_\nu$ with three texture zeros in the (3+1) model, assuming all the neutrino states are of Majorana type. With the mass hierarchy between the active and sterile neutrinos as well as the smallness of the reactor neutrino mixing and active-sterile mixing, the analytical expressions can be greatly simplified. We systematically examine all the 120 texture zeros, via both analytical and numerical analysis, and find that the 100 textures with zeros in the fourth column and row of $M_\nu$ can hardly be compatible with experiments; whereas, 19 of the other 20 are feasible. We can foresee that the active neutrino masses are required to be around 0.01~eV and the next generation neutrinoless double beta decay experiments are promising to test these textures. Many other phenomenologically meaningful predictions are also obtained. All the CP conserving (3+1) textures with three zeros are excluded by experiments.
\end{abstract}
\draft
\maketitle

\tableofcontents

\newpage
\section{Introduction}

After observation of the smallest neutrino oscillation by Daya Bay~\cite{theta13-dyb} for the first time at the C.L. of $5\sigma$, and immediately further confirmed by RENO~\cite{theta13-reno}, dozens of solar, atmospheric, reactor, and accelerator-based neutrino experiments have compellingly provided us with solid information on five physical parameters involving three active flavors of massive neutrinos~\cite{pdg2012}. The three mixing angles quantify the difference between the flavor and mass eigenstates of the three generations of neutrinos and the amplitude of neutrino oscillations; the two independent mass squared differences are directly connected to neutrino mass spectrum and crucial to the oscillation frequencies in neutrino experiments. Although our knowledge on neutrinos in the standard model (SM) of particle physics has achieved great progress, many basic properties of these magic particles are unclear and to be explored, e.g., their absolute masses, the mechanism for their tiny masses, and CP violation in the lepton sector.

One puzzle on neutrinos is the long-standing LSND anomaly~\cite{LSND}, which points to the existence of one eV scale neutrino and cannot be accommodated in the three-neutrino paradigm. The anomalous $\bar{\nu}_\nu \rightarrow \bar{\nu}_e$ in LSND motivated the MiniBooNE experiment~\cite{MiniBooNE}, whose latest low energy data support the LSND result, although the neutrino and antineutrino mode of the MiniBooNE experiment are incompatible to a large extent~\cite{MiniBooNE2012}. Furthermore, recalculation of the reactor $\bar{\nu}_e$ flux indicates the anomalous $\bar{\nu}_e$ disappearance~\cite{sn-hint-reactor}, i.e., the so-called ``reactor anomaly''. The Gallium radioactive source experiments GALLEX and SAGE~\cite{sn-hint-Gallium} have also reported hints of $\nu_e$ disappearance, corresponding to the eV scale neutrino oscillation. In addition, cosmological data on cosmological microwave background, large scale structure, and the Hubble constant imply extra light degrees of freedom~\cite{sn-whitepaper}, and light extra neutrinos beyond the three in the SM are convincingly the most natural candidate. All these hints of extra neutrinos, however, are restricted by, e.g., the ICARUS\cite{ICARUS} and $\nu_e - ^{12}\!{\rm C}$ scattering~\cite{nueC} data.

Due to the stringent limit on the $Z$ boson decay width~\cite{Zdecaywidth}, the extra light neutrinos cannot directly participate in the weak interaction and is consequently sterile. The simplest extension of the lepton sector is the addition of one sterile flavor to the three active ones, i.e., the resultant (3+1) neutrino model. In this paper we concentrate on the simplest extension, although introducing more sterile states may reconcile comparatively better the data sets on sterile neutrinos and obtain a more preferable goodness-of-fit~\cite{sn-fit-Conrad,sn-fit-Maltoni}.

Neutrino mass matrix $M_\nu$ with texture zeros, i.e., vanishing elements, is an attractive framework to phenomenologically study neutrinos and the connection of neutrino physics to SM and new physics beyond SM. Texture zero structures imply the underlying flavor symmetry or mechanism~\cite{flavor-symmetry}, e.g., the $Z_n$ symmetry or Froggatt-Nielsen mechanism~\cite{FN-mechanism}. Furthermore, with vanishing elements in $M_\nu$, the physical parameters on neutrino mass, mixing and CP violation are always connected to each other. The framework of texture zeros have been intensively studied in the three-neutrino frame~\cite{FGM,Xing,zero,2zero,zero-Dirac,zero-minor,zero-permutation,zero-Cabibbo} and it can likewise apply to the (3+1) model with an extra sterile neutrino~\cite{2zero-sterile}. In the (3+1) model, if all the four neutrino states are of Majorana type, the neutrino mass matrix $M_\nu$ has ten independent elements,
\begin{equation}
\label{matrix:Mnu}
M_\nu = \left( \begin{array}{cccc}
m_{ee} & m_{e\mu} & m_{e\tau} & m_{es} \\
m_{e\mu} & m_{ \mu \mu} & m_{\mu \tau}& m_{\mu s} \\
m_{e\tau} & m_{\mu\tau} & m_{\tau \tau}& m_{\tau s}\\
m_{es} & m_{\mu s} & m_{\tau s}  & m_{ss} \end{array} \right) \,,
\end{equation}
compared to six in the three-neutrino scheme. Ghosh~\emph{et al.}~\cite{2zero-sterile} has recently investigated the (3+1) model with two zeros, analyzing systematically all the ${\rm C}_{10}^2 = 45$ distinct textures. They found that the 15 textures with both the two zeros in the active sector (the first three rows and columns) of $M_\nu$ are compatible with current neutrino data, whereas the 30 with zero(s) in the sterile sector (the fourth row and column) are excluded by experiments mainly due to the mass hierarchy between the active and sterile neutrinos.\footnote{Actually their analysis is inadequate and not all the 30 structures with zero(s) in the sterile sector are definitely excluded. If $m_{\tau s}$, as well as one matrix element in the active sector, vanishes, one resultant possibility is that the mixing angle $\theta_{34}$ is close to zero. As the short baseline experiments are not sensitive to the tauon flavor, the vanishing of $\theta_{34}$ is allowed~\cite{sn-fit-theta34} and it is yet immature to assert that such textures are excluded. See the following analytical and numerical discussions for details.}

In the presence of CP violation phases, one zero in $M_\nu$ generally imposes two correlation conditions on the parameters on neutrino masses, mixing angles and CP violating phases, i.e., one is related to the vanishing absolute value and another to the vanishing argument. In general, more zeros means more stringent constraints on neutrino parameters and consequently more accurate predictions. In this work we focuses on $4\times4$ neutrino mass matrices with three vanishing elements, assuming all the four neutrino states are of Majorana type. As a comparison, none of the 20 $3\times3$ mass matrices with three zeros are compatible with current data~\cite{FGM,Xing}. Adding one sterile neutrino, the number of possible textures with three zeros sum up to ${\rm C}_{10}^3 = 120$ and this work aims to investigate the feasibility of three zero textures in the (3+1) scheme and explore their phenomenological implications.

Some elements of the lepton mixing matrix are more lengthy in the (3+1) model than in the three neutrino scheme~\cite{parametrization} and resultantly most of the elements of $M_\nu$ become even more lengthy. However, due to the large mass hierarchy between the sterile and active neutrinos $m_4 \gg m_{1,\,2,\,3}$ and the smallness of the angles governing reactor mixing~\cite{theta13-dyb,theta13-reno} and active-sterile mixing (ASM)~\cite{sn-fit-Conrad,sn-fit-Conrad2,sn-fit-Giunti,sn-fit-Giunti2,sn-fit-Maltoni}, the vanishing conditions $m_{\alpha\beta}=0$ ($\alpha,\,\beta = e,\,\mu,\,\tau,\,s$) can be unbelievably simplified and we arrive at clean expressions for the three ASM angles. Based on these expressions we predict the restricted ranges of the ASM angles and other phenomenological observables, e.g., the effective Majorana electron neutrino mass $\langle m_{ee} \rangle$ in neutrinoless double beta decay ($0\nu\beta\beta$).

We also numerically analyse these textures and study their phenomenological implications. We apply experimental constraints on the five active neutrino oscillation parameters, making use of the $3\sigma$ ranges from Ref.~\cite{nu-fit}, and examine which textures can accommodate the eV scale sterile neutrino mass and $\sim$0.1 ASMs implied by short baseline (SBL) experiments. The sterile neutrino fits~\cite{sn-fit-Conrad,sn-fit-Conrad2,sn-fit-Giunti,sn-fit-Giunti2,sn-fit-Maltoni} are somehow different from each other due to variations of the experimental data and calculation procedure, and we assign the fit in Ref.~\cite{sn-fit-Giunti} as the criterion on the three parameters: LSND mass square difference $\Delta m^2_{41}$, the ASM angle $\theta_{14}$ relating $\nu_e$ and $\nu_s$, and $\theta_{24}$ relating $\nu_\mu$ and $\nu_s$. Specifically, as cosmological data prefer lighter neutrino states, we use the $3\sigma$ parameter regions in the $\Delta m^2_{41} - \sin^22\theta_{ee}$ and $\Delta m^2_{41} - \sin^22\theta_{\mu\mu}$ planes with $\sim {\rm eV}^2$ $\Delta m^2_{41}$ in Fig.~1 of Ref.~\cite{sn-fit-Giunti}, where the effective mixing angles are defined as
\begin{eqnarray}
\sin^22\theta_{\alpha\alpha} \equiv 4|U_{\alpha4}|^2 (1-|U_{\alpha4}|^2) \,,
\end{eqnarray}
with $\alpha = e,\,\mu$. SBL experiments are not sensitive to the neutrino of tauon flavor, then the constraint on the third ASM angle $\theta_{34}$ is comparatively rather loose, $0^\circ < \theta_{34} < 30^\circ$~\cite{sn-fit-theta34}. In our numerical calculation, all the randomly generated parameters constrained by experiments are evenly distributed in their $3\sigma$ ranges. Other parameters, e.g., the CP violating phases $\delta_{13}$, have not been determined at the $3\sigma$ C.L. If they are randomly generated, we assume they distribute evenly in their logically allowed ranges, e.g., $\delta_{13} \in [0, 2\pi]$.

The paper is organized as follows: The 120 textures with three zeros in the (3+1) model are systematically classified and labeled in the Appendix. The formulas and experimental constraints on the neutrino parameters are briefly exposited in Sec.~\ref{sec:notation}. Section~\ref{sec:analytical} is devoted to analytical analysis of the textures and Sec.~\ref{sec:calculation} to the numerical discussions. We briefly comment on the textures without CP violation in Sec.~\ref{sec:cpconservation} and end up with the conclusion in Sec.~\ref{sec:conclusion}.

\section{Formulas and experimental constraints}
\label{sec:notation}

Before demonstration of the analytical and numerical calculations, it is necessary to clarify our notations of the neutrino parameters and briefly explain relevant formulas, which is the aim of this section. We also sketch the current experimental constraints on the neutrino parameters.

\subsection{Notations  and formulas}

Assuming the charged lepton mass matrix is diagonal and positive-definite, the $4\times4$ Majorana neutrino mass matrix $M_{\nu}$ can be expressed as, in the flavor basis,
\begin{equation}
\label{Mnu}
M_{\nu}=VM_{\nu}^{\rm diag}V^{T} \,,
\end{equation}
where $M_{\nu}^{\rm diag} = {\rm diag}\{ m_1, m_2, m_3, m_4 \}$ is the diagonal neutrino matrix in the mass basis, and $V$ is the $4\times4$ PMNS lepton mixing matrix.
In general, $V$ can split into two parts,
\begin{eqnarray}
\label{matrix:V}
  V=UP \,.
\end{eqnarray}
$U$ is the lepton mixing matrix for Dirac neutrinos, containing the solar mixing angle $\theta_{12}$, atmospheric mixing angle $\theta_{23}$, reactor mixing angle $\theta_{13}$, three ASM angles $\theta_{14}$, $\theta_{24}$, and $\theta_{34}$, and three Dirac CP violating phases. If we parametrize $U$ explicitly as~\cite{lsn-ph}
\begin{eqnarray}
\label{matrix:U}
&& U = ({R_{34}}\tilde{R}_{24}\tilde{R}_{14})\;(R_{23}\tilde{R}_{13})\;R_{12} \,,
\end{eqnarray}
the three Dirac phases are $\delta_{13}$, $\delta_{14}$, and $\delta_{24}$, where $R_{ij}/\tilde{R}_{ij}$ is the rotation matrix in the \textit{ij} flavor space, e.g.,
\begin{subequations}
\begin{align}
&R_{34} = \left( \begin{array}{cccc}
    1~ &~0 & 0 & 0 \\  0~ &~ 0 & 1 & 0 \\ 0~ & ~0 & c_{34}& s_{34} \\0 ~& ~0 & -s_{34} & c_{34}
  \end{array} \right) \,, \\
&\tilde{R}_{14} = \left( \begin{array}{cccc}
    c_{14}~ & ~0 &~ ~0 &~s_{14}e^{-i \delta_{14}} \\ 0 ~ & ~ 1&~~ 0 & 0 \\ 0 ~& ~0 &~~ 1 & 0 \\-s_{14}e^{i \delta_{14}}  & ~ 0& ~~0 &c_{14}
  \end{array} \right) \,,
\end{align}
\end{subequations}
with $s_{ij}\equiv\sin \theta_{ij}$ and $c_{ij}\equiv\cos \theta_{ij}$. The matrix $P$ in Eq.~(\ref{matrix:V}) is a diagonal Majorana phase matrix written as
\begin{eqnarray}
&& P = {\rm diag}\{ 1, e^{i\alpha/2}, e^{i\beta/2}, e^{i\gamma/2} \} \,.
\end{eqnarray}
The factors $1/2$'s are introduced as a consequence of the mixing matrix $V$ appearing twice in Eq.~(\ref{Mnu}).

Noticing that $M_{\nu}^{\rm diag}$ is sandwiched by the diagonal phase matrix $P$, we define
\begin{subequations}
\label{lambda}
\begin{align}
& \lambda_1 \equiv m_1 \,, \\
& \lambda_2 \equiv m_2 e^{i\alpha} \,, \\
& \lambda_3 \equiv m_3 e^{i\beta} \,,  \\
& \lambda_4 \equiv m_4 e^{i\gamma} \,,
\end{align}
\end{subequations}
and the $4\times4$ neutrino mass matrix is slightly simplified to
\begin{equation}
M_{\nu}=U\:{\rm diag}\{ \lambda_1, \lambda_2, \lambda_3, \lambda_4 \}\:U^{T} \,.
\end{equation}
When three independent entries of $M_\nu$ are set to zero,
\begin{subequations}
\label{3zero}
\begin{align}
& M_{\nu,\, ab} = m_{ab} = 0 \,, \\
& M_{\nu,\, cd}= m_{cd}=0 \,, \\
& M_{\nu,\, ef}= m_{ef}=0 \,,
\end{align}
\end{subequations}
where $a,\,b,\,c,\,d,\,e,\,f = e,\,\mu,\,\tau,\,s$, or, explicitly,
\begin{subequations}
\label{lambdaN}
\begin{align}
& \lambda_1U_{a1} U_{b1} + \lambda_2U_{a2} U_{b2} + \lambda_3U_{a3} U_{b3}  + \lambda_4 U_{a4} U_{b4}=0 \,, \\
& \lambda_1U_{c1} U_{d1} + \lambda_2U_{c2} U_{d2} + \lambda_3U_{c3} U_{d3}  + \lambda_4 U_{c4} U_{d4}=0 \,, \\
& \lambda_1U_{e1} U_{f1} + \lambda_2U_{e2} U_{f2} + \lambda_3U_{e3} U_{f3}  + \lambda_4 U_{e4} U_{f4}=0 \,.
\end{align}
\end{subequations}
As we have only three complex equations, less than the number of neutrino flavors in the (3+1) model, we divide $\lambda_{2,\,3,\,4}$ by $\lambda_1\equiv m_1$ and obtain the three following quantities:
\begin{subequations}
\begin{align}
&  \lambda_{21} \equiv \lambda_2/ \lambda_1 = \frac{m_2}{m_1}e^{i\alpha} \,,  \\
&  \lambda_{31} \equiv \lambda_3/ \lambda_1 = \frac{m_3}{m_1}e^{i\beta}  \,,  \\
&  \lambda_{41} \equiv \lambda_4/ \lambda_1 = \frac{m_4}{m_1}e^{i\gamma} \,.
\end{align}
\label{lambdaNN}
\end{subequations}
With three zeros in the $4\times4$ matrix $M_\nu$, based on Eqs.~(\ref{lambdaN}) and (\ref{lambdaNN}), $\lambda_{21,\,31,\,41}$ can be expressed as functions of the matrix elements of $U$,
\begin{subequations}
\begin{eqnarray}
&  \lambda_{21} = \lambda_{21N}/ \lambda_{2341} \,, \\
&  \lambda_{31} = \lambda_{31N}/ \lambda_{2341} \,, \\
&  \lambda_{41} = \lambda_{41N}/ \lambda_{2341} \,,
\end{eqnarray}
\label{lambdaNNN}
\end{subequations}
where
\begin{subequations}
\begin{align}
&  \lambda_{2341} = - \left| \begin{array}{ccc}
                 U_{a2}U_{b2} & U_{a3}U_{b3} & U_{a4}U_{b4} \\
                 U_{c2}U_{d2} & U_{c3}U_{d3} & U_{c4}U_{d4} \\
                 U_{e2}U_{f2} & U_{e3}U_{f3} & U_{e4}U_{f4} \end{array} \right| \,, \\
&  \lambda_{21N} = \left| \begin{array}{ccc}
                 U_{a1}U_{b1} & U_{a3}U_{b3} & U_{a4}U_{b4} \\
                 U_{c1}U_{d1} & U_{c3}U_{d3} & U_{c4}U_{d4} \\
                 U_{e1}U_{f1} & U_{e3}U_{f3} & U_{e4}U_{f4} \end{array} \right| \,, \\
&  \lambda_{31N} = -\left| \begin{array}{ccc}
                 U_{a1}U_{b1} & U_{a2}U_{b2} & U_{a4}U_{b4} \\
                 U_{c1}U_{d1} & U_{c2}U_{d2} & U_{c4}U_{d4} \\
                 U_{e1}U_{f1} & U_{e2}U_{f2} & U_{e4}U_{f4} \end{array} \right| \,, \\
&  \lambda_{41N} =  \left| \begin{array}{ccc}
                 U_{a1}U_{b1} & U_{a2}U_{b2} & U_{a3}U_{b3} \\
                 U_{c1}U_{d1} & U_{c2}U_{d2} & U_{c3}U_{d3} \\
                 U_{e1}U_{f1} & U_{e2}U_{f2} & U_{e3}U_{f3} \end{array} \right| \,.
\end{align}
\end{subequations}
In presence of the three Dirac CP violating phases $\delta_{13}$, $\delta_{14}$ and $\delta_{24}$, the matrix $U$ is generally complex, then, from the definitions of $\lambda_{21,\,31,\,41}$, we can extract three independent neutrino mass ratios
\begin{subequations}
\label{massratio}
\begin{align}
& \frac{m_2}{m_1} = |\lambda_{21}| = |\lambda_{21N}/ \lambda_{2341}| \,, \\
& \frac{m_3}{m_1} = |\lambda_{31}| = |\lambda_{31N}/ \lambda_{2341}| \,, \\
& \frac{m_4}{m_1} = |\lambda_{41}| = |\lambda_{41N}/ \lambda_{2341}| \,,
\end{align}
\end{subequations}
and the three Majorana CP violating phases
\begin{subequations}
\label{Majoranaphase}
\begin{align}
& \alpha = \arg(\lambda_{21})  = \arg(\lambda_{21N}/ \lambda_{2341}) \,, \\
& \beta = \arg(\lambda_{31})   = \arg(\lambda_{31N}/ \lambda_{2341}) \,, \\
& \gamma = \arg(\lambda_{41})  = \arg(\lambda_{41N}/ \lambda_{2341}) \,.
\end{align}
\end{subequations}
Given the three vanishing conditions on $M_\nu$, i.e., Eq.~(\ref{3zero}), the neutrino mass ratios and Majorana phases are related to the six lepton mixing angles and the three Dirac CP violating phases. To further relate these neutrino parameters to neutrino oscillation experiments and utilize the experimental constraints on the solar neutrino mass square difference $\Delta m^2_{21}$, the atmospheric neutrino mass square difference $|\Delta m^2_{31}|$ as well as $\Delta m^2_{41}$, we define the two independent neutrino mass square difference ratios
\begin{subequations}
\label{Rnu}
\begin{align}
& R_{\nu1} \equiv \frac{\Delta m_{21}^2}{|\Delta m_{31}^2|}=\frac{|\lambda_{21}|^2-1}{ \left||\lambda_{31}|^2-1\right|} \,, \\
& R_{\nu2} \equiv \frac{\Delta m_{21}^2}{\Delta m_{41}^2}=\frac{|\lambda_{21}|^2-1}{ |\lambda_{41}|^2-1} \,.
\end{align}
\end{subequations}

\subsection{Experimental constraints on neutrino parameters}

For the convenience of the analytical and numerical calculations below, let us first clearly count the relevant neutrino parameters we are confront with. In presence of the extra sterile flavor, we have altogether 16 physical parameters
\begin{itemize}
  \item
  Six mixing angles between different flavors: three denote the mixing between active flavors, $\theta_{12,\,23,\,13}$, while the other three are relevant to the ASMs, $\theta_{14,\,24,\,34}$.
  \item
  Six CP violating phases: three are of Dirac type, $\delta_{13,\,14,\,24}$, and the other three are of Majorana type, $\alpha$, $\beta$, and $\gamma$.
  \item
  Four neutrino masses: $m_1$, $m_2$, $m_3$, and $m_4$, with the first three predominantly of active flavors, and the last one predominantly of sterile.
\end{itemize}

Currently, the three active neutrino mixing angles and two mass squared differences $\Delta m^2_{21}$ and $|\Delta m^2_{31}|$ can be extracted from neutrino oscillation data. Their corresponding central values and $3\sigma$ ranges are listed in Table~\ref{table:nu-fit}. The Dirac CP violating phase $\delta_{13}$ in the (3+1) model corresponds to the sole Dirac phase $\delta$ in the three-neutrino scheme. Although it can be loosely constrained in global neutrino fits, we treat it as a free parameter due to the low sensitivity of current data to CP violation~\cite{nu-fit,nu-fit2}.
\begin{table}[tp]
  \centering
  \caption{Experimental data on neutrino oscillation parameters~\cite{nu-fit}.}
  \label{table:nu-fit}
  \begin{tabular}{cccc}
  \hline\hline
    Parameter & Hierarchy & Best fit  & 3$\sigma$ range    \\ \hline
    $\Delta m^2_{21}\; [10^{-5}\text{eV}^2]$  & NH/IH & 7.62          & $7.12 - 8.20$ \\ \hline
    $|\Delta m^2_{31}|\; [10^{-3}\text{eV}^2]$       &NH     & 2.55          & $2.31 - 2.74$ \\
                                              &IH     & 2.43          & $2.21 - 2.64$ \\ \hline
    $\theta_{12}$                             & NH/IH & $34.4^\circ$  & $31.3^\circ - 37.5^\circ$ \\ \hline
    $\theta_{23}$                             & NH    & $51.5^\circ$  & $36.9^\circ - 55.6^\circ$ \\
                                              & IH    & $50.8^\circ$  & $37.5^\circ - 54.9^\circ$ \\ \hline
    $\theta_{13}$                             & NH    & $9.0^\circ$   & $7.5^\circ - 10.5^\circ$ \\
                                              & IH    & $9.1^\circ$   & $7.5^\circ - 10.5^\circ$ \\ \hline\hline
  \end{tabular}
\end{table}

We assume that the sterile neutrino is heavier than the three active ones, i.e., the LSND mass squared difference is positive $\Delta m^2_{41} > 0$. In the opposite case $\Delta m^2_{41}<0$ with three heavier active neutrinos, the sum of neutrino masses $\Sigma_m$ will be $\sim3$ eV, which is much less favored than the former case by cosmological observations~\cite{pdg2012}. The sign of $\Delta m^2_{31}$ has not been pinned down and then two mass spectra are allowed: normal hierarchy (NH) with $m_3 > m_{1,\,2}$ and inverted hierarchy (IH) with $m_3 < m_{1,\,2}$. The lightest neutrino mass $m_0$ in the two spectrums are, respectively, $m_0=m_1$ and $m_0=m_3$.

As mentioned in the introduction, the $3\sigma$ bound on $\Delta m^2_{41}$ and $\theta_{14,\,24}$ used in this work is from Fig.~1 of Ref.~\cite{sn-fit-Giunti} and the constraint on $\theta_{34}$ is from Ref.~\cite{sn-fit-theta34}, as presented in Table~\ref{table:sn-fit}. Beware that the upper and lower bounds of $\theta_{14}$ and $\theta_{24}$ are dependent of $\Delta m^2_{41}$ and the bounds given in Table~\ref{table:sn-fit} are only the allowed maximum and minimum values. It is phenomenologically meaningful to investigate whether these three-zero textures are compatible with the eV sterile neutrino mass and small ASM. To this end, $\Delta m^2_{41}$ treated as input parameters, we calculate analytically and numerically predictions of the textures on the three ASM angles and compare them with experimental constraints. Consequently, if the 16 neutrino parameters are roughly classified into the ``know'' and the ``unknown'', we have six ``known'' parameters: the five active neutrino parameters ($\theta_{12,\,23,\,13}$, $\Delta m^2_{21}$, and $|\Delta m^2_{31}|$) and $\Delta m^2_{41}$. The left 10 are otherwise treated as ``unknown'': the three ASM angles, the six CP violating phases, and the lightest neutrino mass $m_0$.
\begin{table}[tp]
  \centering
  \caption{Current constraints on sterile neutrino parameters~\cite{sn-fit-Giunti,sn-fit-theta34}. See text for the comment on the angles $\theta_{14,\,24}$.}
  \label{table:sn-fit}
  \begin{tabular}{ccc}
  \hline\hline
    Parameter & Hierarchy  & 3$\sigma$ range    \\  \hline
    $\Delta m^2_{41} \; [\text{eV}^2]$  & NH/IH    & $0.72 - 2.53$ \\ \hline
    $\theta_{14}$                       & NH/IH    & $6.1^\circ - 15.3^\circ$ \\ \hline
    $\theta_{24}$                       & NH/IH    & $2.8^\circ - 12.5^\circ$ \\ \hline
    $\theta_{34}$                       & NH/IH    & $0^\circ - 30^\circ$ \\ \hline\hline
  \end{tabular}
\end{table}

In general, the three vanishing elements of $M_\nu$ induce six independent relations among the neutrino masses, mixing angles, and CP violating phases, i.e., Eq.~(\ref{massratio}) and (\ref{Majoranaphase}). That is to say, the 10 ``unknown'' neutrino parameters are constrained by six correlation conditions and we are left with four degrees of freedom completely unconstrained. In other words, generally we cannot completely determine the 10 ``unknown'' parameters in the (3+1) model with three zeros. However, according to the analytical and numerical calculations below, we are able to obtain phenomenologically meaningful predictions on neutrino mass spectrum, ASMs and CP violating phases. Furthermore, we can also foresee the constrained ranges of the nonoscillation quantities: the sum of neutrino masses
\begin{equation}
  \Sigma_m = \sum_{i=1}^4 m_i \,,
\end{equation}
the effective electron neutrino mass in tritium beta decay
\begin{equation}
  m_\beta = \sqrt{\sum_{i=1}^4 m_i^2 |V_{ei}|^2} \,,
\end{equation}
and the effective Majorana electron neutrino mass governing $0\nu\beta\beta$:
\begin{equation}
  \langle m_{ee} \rangle = m_{ee} = \left| \sum_{i=1}^4 m_i V_{ei}^2 \right| \,.
\end{equation}

One may notice that if all the three Dirac CP violating phases vanish, according to the relations in Eq.~(\ref{Majoranaphase}), the three Majorana phases are also zero, and the lepton mixing matrix $U$ is real.\footnote{In the absence of CP violation, the quantities defined in Eq.~(\ref{lambda}) are modified to $\lambda_i \equiv m_i$ with $i=1,\,2,\,3,\,4$, and resultantly, the quantities in Eq.~(\ref{lambdaNN}) are reduced to $\lambda_{i1} \equiv m_i/m_1$ with $i=2,\,3,\,4$.} In this simplified case, we are left with only three correlation conditions of form
\begin{equation}
\frac{m_i}{m_1} = \lambda_{i1} = \lambda_{i1N}/ \lambda_{2341} \,,
\end{equation}
with $i=2,\,3,\,4$, and the ten ``unknown'' parameters shrink to four, i.e., the three ASM angles and the lightest neutrino mass $m_0$. Clearly, the correlation conditions are not yet sufficient to completely determine all the neutrino relevant parameters. We will also briefly investigate the CP conserving three-zero textures, via both analytical and numerical analysis, to judge whether they are compatible with current neutrino data.

\section{Analytical analysis}
\label{sec:analytical}

It seems impossible or unrealistic to obtain clean analytical expressions in the (3+1) model, as some of the elements of matrix $U$ is rather lengthy. However, when we take into account the mass hierarchy $m_4 \gg m_{1,\,2,\,3}$ and the smallness of the three ASM angles and $\theta_{13}$, all the lengthy expressions can be sharply shortened and it becomes viable to analytically examine the $4\times4$ textures of $M_\nu$ with three zeros. The 120 textures has been classified and labelled in the Appendix. In this section we first analyze 99 of the 100 type I textures with zeros in the sterile sector and then the 20 of type II with all the three zeros in the active sector. For the texture compatible with experiments, we give explicit expressions and predicted ranges of the ASM angles and investigate the implications for the neutrino spectrum, CP violating phases, sum of neutrino masses and the effective neutrino mass in single and double beta decays.

\subsection{Textures of type~I}

All the 100 type I textures with three zeros in the (3+1) model have sterile relevant vanishing elements, i.e., the vanishing elements in the fourth column and the fourth row. If the fourth diagonal element $m_{ss}$ of $M_\nu$ vanishes (class $\mathcal{E}$),
\begin{equation}
  m_1U_{s1}^2 +m_2U_{s2}^2e^{i\alpha} +m_3U_{s3}^2e^{i\beta} +m_4U_{s4}^2e^{i\gamma} =0 \,.
\end{equation}
As $m_{1,\,2,\,3} \ll m_4$, to satisfy the equation, it is required that $U_{s4} =\cos\theta_{14}\cos\theta_{24}\cos\theta_{34} \ll 1$, i.e., at least one of the three ASM angles be close to $\pi/2$, which is not compatible with the small ASMs implied by SBL experiments, as mentioned in Ref.~\cite{2zero-sterile}. Furthermore, if $\theta_{14} \sim \pi/2$, it yields that
\begin{equation}
\label{mss}
  m_\beta \sim \langle m_{ee}\rangle \sim m_4 \sim {\rm eV} \,.
\end{equation}
The current most stringent limit on $\langle m_{ee}\rangle$ is from the combined result of KamLAND-Zen and EXO-200, $\langle m_{ee}\rangle < (0.12 - 0.25)$~eV at 90\% C.L.~\cite{0nubetabeta-ex}, strongly disfavoring the value given in Eq.~(\ref{mss}).

If one of the nondiagonal entries in the sterile sector is set to zero, i.e., $m_{is}=0$ with $i=e,\,\mu,\,\tau$ (class $\mathcal{F}$),
\begin{equation}
\label{mis}
  m_1U_{i1}U_{s1} +m_2U_{i2}U_{s2}e^{i\alpha} +m_3U_{i3}U_{s3}e^{i\beta} +m_4U_{i4}U_{s4}e^{i\gamma} =0 \,.
\end{equation}
For the same reason of $m_{1,\,2,\,3} \ll m_4$, the equation requires $U_{i4}U_{s4} \ll 1$. As $U_{e4} \sim s_{14}$, $U_{\mu4} \sim c_{14}s_{24}$, $U_{\tau4} = c_{14}c_{24}s_{34}$, and $U_{s4} =c_{14}c_{24}c_{34}$, we arrive at the conclusion that, to satisfy Eq.~(\ref{mis}), at least one of the two following conditions must be satisfied: (1) To have a vanishing entry $m_{is}$ ($i=e,\,\mu$), the corresponding mixing angle between the active flavor $i$ and the sterile one has to be nearly vanishing; or/and (2) At least one of the three ASM angles is close to $\pi/2$. Unfortunately, both of these two extreme conditions are excluded by SBL experiments.

In the discussion above we did not consider the textures with solely $m_{\tau s}$ in the sterile sector, as, different from $\theta_{14,\,24}$, a vanishing $\theta_{34}$ is allowed due to the insensitivity of SBL experiments to $\nu_\tau$. If $\theta_{14,\,24}$ are mildly small, say $\sim 10^\circ$, with the leading order (LO) approximation $m_4 \gg m_{1,\,2,\,3}$ and $\theta_{13,\,14,\,24} \ll 1$, the third ASM angle is estimated from Eq.~(\ref{mis}) to be
\begin{eqnarray}
\theta_{34} \sim \frac{\sqrt{|\Delta m^2_{31}|}}{\sqrt{\Delta m^2_{41}}} \,\theta_{14} \sim 0.4^\circ \,.
\end{eqnarray}
In this case, the textures of class $\mathcal{F}$ with $m_{\tau s}=0$ cannot be completely excluded. However, such solutions are highly disfavored, as the parameter space in this specific region is rather small, which can be verified in the numerical calculation below.

If the textures have both $m_{ss}=0$ and $m_{is}=0$ with $i=e,\,\mu,\,\tau$ (classes $\mathcal{A}$ and $\mathcal{C}$), as for class $\mathcal{E}$, $c_{14}c_{24}c_{34} \ll 1$ are required and all these texture are excluded by SBL experiments. In the texture of class $\mathcal{B}$ there is actually no mixing between the active and sterile sector and this texture is completely of no interest for us. For the textures of class $\mathcal{D}$, analogical to the class $\mathcal{F}$, two of the three ASM angles are required to be rather small or rather large, which are not allowed by SBL experiments. In short, all the (3+1) textures with zeros in the sterile sector are excluded by SBL experiments, except for the textures of class $\mathcal{F}$ with $m_{\tau s}=0$ which, however, are highly disfavored.

\subsection{Textures of type II}

In three-neutrino scheme the $M_\nu$ textures with three zeros are definitely not compatible with the current neutrino oscillation data~\cite{Xing}. When the lepton sector of SM is extended to include one light sterile neutrino, one zero in the active sector of $M_\nu$ reads
\begin{equation}
\label{eq:active}
m_1U_{i1}U_{j1} +m_2U_{i2}U_{j2}e^{i\alpha} +m_3U_{i3}U_{j3}e^{i\beta} +m_4U_{i4}U_{j4}e^{i\gamma} =0 \,,
\end{equation}
with $i,\,j=e,\,\mu,\,\tau$. Compared to the three-neutrino framework, not only the fourth term is new, but the elements of $U$ are tuned by six more neutrino parameters (three ASM angles and three more CP violating phases). All these facts provide the possibility of reconcile the experimental bound on neutrino parameters and the constraints from vanishing mass matrix elements.

Numerically, $m_{1,\,2,\,3} \ll m_{4}$ and $U_{i4}^2 \sim 0.01$ are generally much smaller than $U_{im}U_{jn}$ ($i,\,j=e,\,\mu,\,\tau$, $m,n=1,\,2,\,3$), thus it is possible for the contributions from active neutrinos to cancel out the fourth term in Eq.~(\ref{eq:active}) and obtain vanishing elements in the active sector of $M_\nu$. If the sterile neutrino does mix with the active ones with strength of order 0.1 as the SBL experiments indicate, we can immediately obtain from Eq.~(\ref{eq:active}) that the active neutrino masses are of order $m_4U_{i4}^2 \sim \sqrt{\Delta m^2_{41}} U_{i4}^2 \sim 0.01$ eV and the lightest neutrino mass is of the same order
\begin{eqnarray}
m_0 \sim 0.01\;{\rm eV} \,.
\end{eqnarray}
Actually, this is a direct consequence of the neutrino mass matrix $M_\nu$ in presence of the sterile neutrino. To be more transparent, let us consider the simplified $2\times2$ version of $M_\nu$ with only the first generation of active neutrino,
\begin{eqnarray}
M_\nu = \left( \begin{array}{cc}  0 & m_{es} \\ m_{es} & m_{ss}  \end{array} \right) \,.
\end{eqnarray}
Due to the large active-sterile mass splitting, we have set $m_{ee}=0$ and are left with a mini-seesaw mechanism at the eV scale~\cite{seesaw-eV}. The ASM is expected to be $\sim m_{es}/m_{ss}\sim0.1$; thus, with an eV scale sterile neutrino $m_{ss}\sim1$~eV, we naturally obtain the active neutrinos are of mass $\sim m_{es}^2/m_{ss}\sim0.01$~eV. This meaningful prediction are not sensitive to the details of the active sector of $M_\mu$ (this part can always be ignored when we apply the seesaw formula) and it is a generic feature for $4\times4$ $M_\nu$ textures with zeros in the \emph{active} sector, regardless of the number of zeros~\cite{2zero-sterile}.
One phenomenological consequence of the prediction on neutrino spectrum is that models yielding a massless state or large mass hierarchy in the active neutrino sector are not compatible with the (3+1) scenarios with texture zeros.

Applying the approximations $m_{1,\,2,\,3} \ll m_{4}$ and $\theta_{13,\,14,\,24,\,34} \ll 1$, the six independent vanishing elements in the active sector of $M_\nu$ are simplified to, at LO,
\begin{subequations}
\label{eqn:II}
\begin{align}
m_{ee} &: \;  m_4 s_{14}^2 e^{-i(2\delta_{14}-\gamma)}
    +\Big( m_1 c_{12}^2
    +m_2 s_{12}^2 e^{i\alpha} \Big) =0 \,, \label{eqn:ee}\\
m_{\mu\mu} &:\;  m_4 s_{24}^2 e^{-i(2\delta_{24}-\gamma)}
    + \Big( m_1 s_{12}^2 c_{23}^2
    +m_2 c_{12}^2 c_{23}^2 e^{i\alpha}
    +m_3 s_{23}^2 e^{i\beta} \Big) =0 \,, \label{eqn:mumu}\\
m_{\tau\tau} &:\;   m_4 s_{34}^2 e^{i \gamma }
   + \Big( m_1 s_{12}^2 s_{23}^2
   +m_2 c_{12}^2 s_{23}^2 e^{i \alpha }
   +m_3 c_{23}^2 e^{i \beta } \Big) =0 \,, \label{eqn:tautau}\\
m_{e\mu} &:\;  m_4 s_{14} s_{24} e^{-i (\delta _{14}+\delta _{24}-\gamma)}
   - \Big( (m_1-m_2e^{i \alpha}) s_{12}c_{12}c_{23} -m_3 s_{23}s_{13}e^{-i (\delta _{13}-\beta)} \Big) =0 \,, \label{eqn:emu} \\
m_{e\tau} &:\;  m_4 s_{14} s_{34} e^{-i(\delta _{14}-\gamma)}
   + \Big( (m_1-m_2e^{i\alpha}) s_{12}c_{12}s_{23}
   +m_3 c_{23} s_{13} e^{-i(\delta _{13}-\beta)} \Big)  =0 \,, \label{eqn:etau} \\
m_{\mu\tau} &:\;  m_4 s_{24} s_{34} e^{-i(\delta_{24}-\gamma)}
   -s_{23}c_{23} \Big( m_1 s_{12}^2
    +m_2 c_{12}^2  e^{i\alpha}
    -m_3  e^{i\beta} \Big) =0 \,. \label{eqn:mutau}
\end{align}
\end{subequations}
We emphasize again that these explicit equations strongly indicate the feasibility of the (3+1) model with zeros in the active sector: Assuming $m_0\sim0.01$~eV, Eq.~(\ref{eqn:II}) implies that the ASM angles
\begin{equation}
s_{i4} \sim
\left[ \frac{\sqrt{\Delta m^2_{21}}}{\sqrt{\Delta m^2_{41}}} \right]^{1/2} \;\; \text{or} \;\;
\left[ \frac{\sqrt{|\Delta m^2_{31}|}}{\sqrt{\Delta m^2_{41}}} \right]^{1/2} \quad
\sim (0.1 - 0.2) \,, \quad (i=1,\,2,\,3)
\end{equation}
coinciding with the hints of SBL experiments. In addition, a few comments on the approximate expressions follow:
\begin{itemize}
  \item
  $U_{e4}\sim s_{14}$, thus after approximation the vanishing condition $m_{ee}=0$ is reduced to the LO expression for $s_{14}^2$, i.e., (\ref{eqn:ee}). Similarly, by means of $U_{\mu4}\sim s_{24}$ and $U_{\tau4}\sim s_{34}$, the two vanishing diagonal elements $m_{\mu\mu}=m_{\tau\tau}=0$ are simplified to the equations for the angles $\theta_{24,\,34}$ [(\ref{eqn:mumu}) and (\ref{eqn:tautau})], while the three nondiagonal vanishing elements $m_{e\mu}=m_{e\tau}=m_{\mu\tau}=0$ lead to the three equations (\ref{eqn:emu}), (\ref{eqn:etau}), and (\ref{eqn:mutau}) of the cross terms.
  \item
  In Eqs.~(\ref{eqn:emu}) and (\ref{eqn:etau}) the terms of form $m_3s_{13}$ are reserved, because for the neutrino spectrum of NH their magnitudes are of the same order of the $m_{1,\,2}$ terms, whereas in the other four equations of (\ref{eqn:II}), the LO terms in $\theta_{13}$ are of form $s_{13}^2$ and are neglected.
  \item
  Comparing the Eqs.~(\ref{eqn:mumu}) and (\ref{eqn:tautau}) as well as (\ref{eqn:emu}) and (\ref{eqn:etau}), these equations exhibit the approximate $\mu-\tau$ symmetry~\cite{zero-permutation}. For example, the textures $M_{\nu}^{(\tau)}$ with $m_{\tau\tau}=0$ can be approximately obtained by acting the $\mu-\tau$ permutation matrix
  \begin{eqnarray}
  P_{\mu\tau} =
  \left( \begin{array}{cccc} 0 & 0 & 0 & 0 \\ 0  & 0 & 1 & 0 \\ 0 & 1 & 0 & 0 \\ 0 & 0 & 0 & 0  \end{array} \right)
  \end{eqnarray}
  on the matrices $M_\nu^{(\mu)}$ with $m_{\mu\mu}=0$~\cite{2zero-sterile}:
  \begin{eqnarray}
  M_{\nu}^{(\tau)} = P_{\mu\tau} M_{\nu}^{(\mu)} P_{\mu\tau}^T \,.
  \end{eqnarray}
  This approximate symmetry will be more transparent if we write explicitly down the analytical expressions of the ASM angles, as shown below.
\end{itemize}

All the discussions of phenomenological implications of the 20 type II texture are based on the equations in (\ref{eqn:II}). With the six equations, we can easily obtain clear expressions of the quantities $\lambda_{21,\,31,\,41}$ and then immediately the neutrino mass ratios and the Majorana phases. However, in the (3+1) model it is more attractive to investigate directly the ASM angles and relevant phenomenologies. We denote the six vanishing conditions of Eq.~(\ref{eqn:II}) simply as
\begin{subequations}
\label{eqn:142434}
\begin{align}
s_{14} &= a \,, \label{eqn:s14} \\
s_{24} &= b \,, \label{eqn:s24} \\
s_{34} &= c \,, \label{eqn:s34} \\
s_{14}s_{24} &= d^2 \,, \label{eqn:s1424} \\
s_{14}s_{34} &= e^2 \,, \label{eqn:s1434} \\
s_{24}s_{34} &= f^2 \,, \label{eqn:s2434}
\end{align}
\end{subequations}
where
\begin{subequations}
\label{eqn:abcdef}
\begin{align}
  a  &\equiv \frac{1}{\sqrt{m_4}}
  \left[ m_1^2 c_{12}^4 +m_2^2 s_{12}^4 +2m_1 m_2 s_{12}^2 c_{12}^2 \cos\alpha \right]^{1/4} \,, \label{eqn:a} \\
  b  &\equiv \frac{1}{\sqrt{m_4}}
  \left[ m_1^2s_{12}^4c_{23}^4 +m_2^2c_{12}^4c_{23}^4 +m_3^2s_{23}^4
        +2 m_1 m_2 s_{12}^2c_{12}^2c_{23}^4 \cos\alpha  \right. \nonumber \\
  & \qquad\qquad \left.  +2m_1m_3 s_{12}^2s_{23}^2c_{23}^2 \cos\beta + 2 m_2 m_3 c_{12}^2s_{23}^2c_{23}^2 \cos(\alpha-\beta)  \right]^{1/4} \,, \label{eqn:b} \\
  c  &\equiv \frac{1}{\sqrt{m_4}}
  \left[ m_1^2 s_{12}^4 s_{23}^4
        +m_2^2 c_{12}^4 s_{23}^4
        +m_3^2 c_{23}^4
        +2 m_1 m_2 s_{12}^2 c_{12}^2 s_{23}^4 \cos\alpha \right. \nonumber \\
  & \qquad\qquad  \left.
        +2 m_1 m_3 s_{12}^2 s_{23}^2 c_{23}^2 \cos\beta
        +2 m_2 m_3 c_{12}^2 s_{23}^2 c_{23}^2 \cos(\alpha-\beta)
         \right]^{1/4} \,, \label{eqn:c} \\
  d  &\equiv \frac{1}{\sqrt{m_4}}
  \left[ (m_1^2+m_2^2) s_{12}^2 c_{12}^2 c_{23}^2
  +m_3^2 s_{23}^2 s_{13}^2
  -2 m_1 m_2 s_{12}^2 c_{12}^2 c_{23}^2 \cos\alpha \right. \nonumber \\
  & \qquad\qquad \left. -2 s_{12} c_{12} s_{23} c_{23} s_{13} ( m_1 m_3 \cos(\beta-\delta_{13})- m_2 m_3 \cos(\alpha-\beta+\delta_{13})) \right]^{1/4} \,, \label{eqn:d} \\
  e  &\equiv \frac{1}{\sqrt{m_4}}
  \left[ (m_1^2+m_2^2) s_{12}^2 c_{12}^2 s_{23}^2
        + m_3^2 c_{23}^2 s_{13}^2
        -2 m_1 m_2 s_{12}^2 c_{12}^2 s_{23}^2 \cos\alpha \right. \nonumber \\
  &\qquad\qquad \left.  +2  s_{12} c_{12} s_{23} c_{23} s_{13} ( m_1 m_3 \cos \left(\beta-\delta_{13}\right)
                        - m_2 m_3  \cos \left(\alpha-\beta +\delta_{13}\right)) \right]^{1/4} \,, \label{eqn:e} \\
  f  &\equiv \frac{1}{\sqrt{m_4}}
  \left[ s_{23}^2 c_{23}^2 ( m_1^2 s_{12}^4   +m_2^2 c_{12}^4   +m_3^2
                           +2 m_1 m_2 s_{12}^2 c_{12}^2 \cos\alpha \right. \nonumber \\
  & \qquad\qquad  \left.    -2 m_1 m_3 s_{12}^2 \cos\beta
                           -2 m_2 m_3 c_{12}^2 \cos(\alpha -\beta))   \right]^{1/4} \,. \label{eqn:f}
\end{align}
\end{subequations}
The expressions in Eq.~(\ref{eqn:abcdef}) and then the explicit formulas for the ASM angles (\ref{eqn:142434}) can be viewed as expressions of the lightest neutrino mass $m_0$, with, for the NH spectrum
\begin{eqnarray}
\begin{array}{l}
m_1=m_0 \,,\\
m_2=\sqrt{m^2_0+\Delta m^2_{21}} \,, \\
m_3=\sqrt{m^2_0+\Delta m^2_{31}} \,, \\
m_4=\sqrt{m^2_0+\Delta m^2_{41}} \,.
\end{array}
\end{eqnarray}
or for IH
\begin{eqnarray}
\begin{array}{l}
m_1=\sqrt{m^2_0+\Delta m^2_{31}}\,,\\
m_2=\sqrt{m^2_0+\Delta m^2_{21}+\Delta m^2_{31}}\,, \\
m_3=m_0\,, \\
m_4=\sqrt{m^2_0+\Delta m^2_{31}+\Delta m^2_{41}}\,.
\end{array}
\end{eqnarray}
In the case of IH, as $m_3=m_0\sim0.01$~eV and consequently the terms $m_{3}s_{13}$ are much smaller than the contribution from $m_{1,\,2}$, the $d$ and $e$ in Eq.~(\ref{eqn:abcdef}) can be further simplified to
\begin{subequations}
\begin{align}
d  &= \frac{1}{\sqrt{m_4}}
\left[ (m_1^2+m_2^2) s_{12}^2 c_{12}^2 c_{23}^2  -2 m_1 m_2 s_{12}^2 c_{12}^2 c_{23}^2 \cos\alpha  \right]^{1/4} \,, \label{eqn:d2} \\
e  &= \frac{1}{\sqrt{m_4}}
\left[ (m_1^2+m_2^2) s_{12}^2 c_{12}^2 s_{23}^2  -2 m_1 m_2 s_{12}^2 c_{12}^2 s_{23}^2 \cos\alpha  \right]^{1/4} \,. \label{eqn:e2}
\end{align}
\end{subequations}
With the four analytical expressions (\ref{eqn:b})-(\ref{eqn:e}) for the ASM angles $s_{24}$, $s_{34}$, $s_{14}s_{24}$, and $s_{14}s_{34}$, the approximate $\mu-\tau$ symmetry is clearer than in Eq.~(\ref{eqn:II}),
\begin{equation}
\label{eqn:mutausymmetry}
\begin{array}{lll}
\theta_{12} \rightarrow \theta_{12}\,, &
\theta_{13} \rightarrow \theta_{13}\,, &
\theta_{23} \rightarrow \frac{\pi}{2}-\theta_{12}\,, \\
\alpha \rightarrow \alpha \,, &
\beta \rightarrow \beta \,, &
\delta_{13} \rightarrow \delta_{13}-\pi \,,
\end{array}
\end{equation}
similar to the case for the schemes with three flavors of neutrinos~\cite{zero-permutation}. It should be reminded that the simple transitions (\ref{eqn:mutausymmetry}) are obtained based on the assumptions of small reactor mixing and ASMs $\theta_{13,\,14,\,24,\,34}\ll1$; the exact formula is a bit more complex, as presented in Ref.~\cite{2zero-sterile}. For the 20 type II textures we now begin to analytically analyze, the approximate $\mu-\tau$ symmetry relates the textures ${\bf B_1}-{\bf B_3}$, ${\bf B_2}-{\bf B_4}$, ${\bf B_5}-{\bf B_6}$, ${\bf C_1}-{\bf C_2}$, ${\bf D_1}-{\bf D_2}$, ${\bf D_3}-{\bf D_5}$, ${\bf D_4}-{\bf D_6}$, and ${\bf E_2}-{\bf E_3}$. In our analytical analysis, we consider all the 20 textures to check whether the approximate symmetry holds, as we expect, for the eight pairs of textures.
\begin{itemize}
  \item
  Texture ${\bf A}$

  The first three diagonal elements of $M_\nu$ vanishes and this texture involve the first three expressions of the ASM angles in Eq.~(\ref{eqn:142434}). Assuming $m_0=0.01$ eV, with the $3\sigma$ ranges for the neutrino mass squared differences $\Delta m^2_{21,\,31,\,41}$ and th mixing angles $\theta_{12,\,23,\,13}$ given in Tables~\ref{table:nu-fit} and {\ref{table:sn-fit}} as well as $\alpha,\,\beta\in[0,\,2\pi]$, one can easily calculate the allowed ranges of the three ASM angles for both the two neutrino spectrums of NH and IH,
  \begin{equation}
  \label{eqn:A-prediction}
  \begin{array}{lcclc}
  1.6^\circ < \theta_{14} < 6.6^\circ & (\mbox{NH}) & \quad\mbox{ or }\quad &
  5.0^\circ < \theta_{14} < 14.4^\circ  & (\mbox{IH}) \,, \\
  4.5^\circ < \theta_{24} < 12.6^\circ & (\mbox{NH}) & \quad\mbox{ or }\quad &
  0 < \theta_{24} < 12.0^\circ  & (\mbox{IH}) \,, \\
  3.9^\circ < \theta_{34} < 12.3^\circ & (\mbox{NH}) & \quad\mbox{ or }\quad &
  0 < \theta_{34} < 12.3^\circ  & (\mbox{IH}) \,.
  \end{array}
  \end{equation}
  The predicted ranges substantially overlap with the ranges determined by SBL experiments, which reveals that both the NH and IH spectrums are allowed in the texture ${\bf A}$. The similar predictions on the angles $\theta_{24,\,34}$ indicates that ${\bf A}$ is approximately self-symmetric under the $\mu-\tau$ permutation $P_{\mu\tau}$. The three textures ${\bf C_3}$, ${\bf E_1}$, and ${\bf F}$ are also self-symmetric under the approximate $\mu-\tau$ permutation, as shown in Tables~\ref{table:asm-exp} and \ref{table:asm-num} below. In Eq.~(\ref{eqn:A-prediction}), the lower limits of the angles $\theta_{24,\,34}$ for IH vanish, because in Eqs.~(\ref{eqn:b}) and (\ref{eqn:c}) it is possible for the contributions from the three light states $m_{1,\,2,\,3}$ to completely cancel out each other. In addition, Eqs.~(\ref{eqn:a})---(\ref{eqn:c}) imply that it is easier to produce vanishing elements ($m_{ee,\,\mu\mu,\,\tau\tau}=0$) if the Majorana CP violating phase $\alpha\sim\pi$  than $\alpha\sim0$, especially for the spectrum of IH. The reason is that with smaller values of the ASM angles, it is more likely for the light states $m_{1,\,2,\,3}$ in Eq.~(\ref{eqn:II}) to cancel out the contribution from the sterile state $m_4$. Although the Majorana phases cannot be observed in neutrino oscillation experiments, they are relevant to the effective neutrino mass in $0\nu\beta\beta$~\cite{0nubetabeta-review}.  We now skip over the discussions of the implications on the effective neutrino mass in single and double beta decays; at the end of this subsection we will collectively analyze these phenomenological implications for all the type II textures.

  \item
  Class ${\bf B}$

  Let us first take the texture ${\bf B_1}$ as an explicit example, in which $m_{e\tau}=0$ in addition to $m_{ee}=m_{\mu\mu}=0$. The LO expressions for $s_{14}$ and $s_{24}$ are the same as for texture {\bf A}, and $s_{13}$ can be solved from Eq.~(\ref{eqn:e}), $s_{34}=e^2/s_{14}=e^2/a$. Assuming again $m_0=0.01$~eV as done for texture ${\bf A}$, we can immediately obtain the predicted ranges of the ASM angles:
  \begin{equation}
  \label{eqn:B1-num}
  \begin{array}{lcclc}
  1.6^\circ < \theta_{14} < 6.6^\circ & (\mbox{NH}) & \quad\mbox{ or }\quad &
  5.0^\circ < \theta_{14} < 14.4^\circ  & (\mbox{IH}) \,, \\
  4.5^\circ < \theta_{24} < 12.6^\circ & (\mbox{NH}) & \quad\mbox{ or }\quad &
  0 < \theta_{24} < 12.0^\circ  & (\mbox{IH}) \,, \\
  0 < \theta_{34} < 26.6^\circ & (\mbox{NH}) & \quad\mbox{ or }\quad &
  0 < \theta_{34} < 23.7^\circ  & (\mbox{IH}) \,,
  \end{array}
  \end{equation}
  which are also compatible with the SBL constraints. Due to the minus sign in Eq.~(\ref{eqn:e2}) for the IH spectrum, when the Majoran phase $\alpha\sim\pi$, the angle $\theta_{34}$ tends to take larger values than with the assumptions of $\alpha\sim0$ (although the other ASM angles $\theta_{14,\,24}$ are likely to be smaller), thus the IH spectrum is less favored than NH.\footnote{As explained for the texture ${\bf A}$, with larger values of $\theta_{34}$, it is harder to produce a vanishing $m_{e\tau}$. Although the SBL constraint on $\theta_{34}$ is less severe than $\theta_{14,\,24}$, it is completely irrelevant to the discussion here.} Due to the same reason of preference for small ASMs, the Majorana phase $\alpha\sim \pi$ is much more favored than $\alpha\sim0$.

  The analysis above apply analogically to the textures ${\bf B_2}-{\bf B_6}$. The compact analytical expressions and the resultant numerical ranges of the three ASM angles in these textures are listed, respectively, in Tables~\ref{table:asm-exp} and \ref{table:asm-num}. In both tables, the textures with the same superscripts are related by the approximate $\mu-\tau$ symmetry, under which the analytical expressions simply transform as
  \begin{eqnarray}
  b \longleftrightarrow c \,, \qquad
  d \longleftrightarrow e \,.
  \end{eqnarray}
  The texture ${\bf B_3}$ is the partner of ${\bf B_1}$ under the $\mu-\tau$ symmetry. Comparing their analytical formula and numerical predictions in Tables~\ref{table:asm-exp} and \ref{table:asm-num}, it is clear that the $\mu-\tau$ symmetry approximately holds, as we expect. These two textures share many similar features: ${\bf B_3}$ also prefers much for NH and $\alpha\sim\pi$.

  Similarly, it is easy to check that the approximate $\mu-\tau$ symmetry holds true for the two other pairs, i.e., ${\bf B_2}-{\bf B_4}$ and ${\bf B_5}-{\bf B_6}$, and we can predict that both NH and IH are compatible with the four structures. With regard to the CP violating phases, the textures ${\bf B_{2,\,4}}$ prefer the values of $\alpha$ around $\pi$. When it comes to ${\bf B_{5,\,6}}$, the situation becomes a bit more complex. In the two textures, both the two vanishing elements $m_{\mu\mu,\,\tau\tau}=0$ involve the two Majorana phases $\alpha$ and $\beta$ at LO, as shown in Eq.~(\ref{eqn:b}) and (\ref{eqn:c}). Among the four approximations of the two phases
  \begin{eqnarray}
  \alpha\sim0\;\&\;\beta\sim0 \,,\quad
  \alpha\sim\pi\;\&\;\beta\sim\pi \,,\quad
  \alpha\sim0\;\&\;\beta\sim\pi \,,\quad
  \alpha\sim\pi\;\&\;\beta\sim0 \,,
  \end{eqnarray}
  for the NH spectrum, small ASMs are most acceptable when the third approximations is satisfied, in which case the contributions from the two states $m_{1,\,2}$ interfere destructively with the third state $m_3$. For IH, we do not have such clear predictions.
  \begin{table}[tp]
  \centering
  \caption{LO expressions of the three ASM angles for the 20 textures of type II. The angle $\theta_{34}$ in the texture ${\bf C}_1$, $\theta_{24}$ in ${\bf C}_2$, and $\theta_{14}$ in ${\bf C}_3$ are unconstrained. The textures with the same superscripts are related by the approximate $\mu-\tau$ symmetry. See text for details.}
  \label{table:asm-exp}
  \begin{tabular}{llll |c llll}
  \hline\hline
  Texture & $s_{14}$ & $s_{24}$ & $s_{34}$          && Texture                &  $s_{14}$ & $s_{24}$ & $s_{34}$ \\ \hline
  ${\bf A}$           & $a$ & $b$ & $c$             && ${\bf F}$              & $\frac{de}{f}$ & $\frac{df}{e}$ & $\frac{ef}{d}$ \\ \hline\hline
  ${\bf B}_1$$^\ast$  & $a$ & $b$ & $\frac{e^2}{a}$ && ${\bf D}_1$$^\P$       & $a$ & $\frac{d^2}{a}$ & $\frac{af^2}{d^2}$       \\ \hline
  ${\bf B}_2$$^\dag$  & $a$ & $b$ & $\frac{f^2}{b}$ && ${\bf D}_2$$^\P$       & $a$ & $\frac{af^2}{e^2}$ & $\frac{e^2}{a}$       \\ \hline
  ${\bf B}_3$$^\ast$  & $a$ & $\frac{d^2}{a}$ & $c$ && ${\bf D}_3$$^\flat$    & $\frac{d^2}{b}$ & $b$ & $\frac{be^2}{d^2}$       \\ \hline
  ${\bf B}_4$$^\dag$  & $a$ & $\frac{f^2}{c}$ & $c$ && ${\bf D}_4$$^\natural$ & $\frac{be^2}{f^2}$ & $b$ & $\frac{f^2}{b}$       \\ \hline
  ${\bf B}_5$$^\ddag$ & $\frac{d^2}{b}$ & $b$ & $c$ && ${\bf D}_5$$^\flat$    & $\frac{e^2}{c}$ & $\frac{cd^2}{e^2}$ & $c$       \\ \hline
  ${\bf B}_6$$^\ddag$ & $\frac{e^2}{c}$ & $b$ & $c$ && ${\bf D}_6$$^\natural$ & $\frac{cd^2}{f^2}$ & $\frac{f^2}{c}$ & $c$       \\ \hline\hline
  ${\bf C}_1$$^\S$    & $a$ & $b$ & $-$             && ${\bf E}_1$            & $a$ & $\frac{d^2}{a}$ & $\frac{e^2}{a}$          \\ \hline
  ${\bf C}_2$$^\S$    & $a$ & $-$ & $c$             && ${\bf E}_2$$^\sharp$   & $\frac{d^2}{b}$ & $b$ & $\frac{f^2}{b}$          \\ \hline
  ${\bf C}_3$         & $-$ & $b$ & $c$             && ${\bf E}_3$$^\sharp$   & $\frac{e^2}{c}$ & $\frac{f^2}{c}$ & $c$          \\ \hline\hline
  \end{tabular}
  \end{table}
  \begin{table}[tp]
  \centering
  \caption{Numerical ranges of the LO analytical expressions given in Table~\ref{table:asm-exp} for the ASM angles in the 20 textures of type II. ``$-$'' denotes that the corresponding angle is unconstrained. The textures with the same superscripts are related by the approximate $\mu-\tau$ symmetry. See text for details.}
  \label{table:asm-num}
  \begin{tabular}{l lll c lll}
  \hline\hline
  & \multicolumn{3}{c}{NH} & & \multicolumn{3}{c}{IH} \\ \cline{2-4} \cline{6-8}
  Texture & $\theta_{14}$ & $\theta_{24}$ & $\theta_{34}$ && $\theta_{14}$ & $\theta_{24}$ & $\theta_{34}$ \\ \hline
  ${\bf A}$              & $[1.6^\circ,\,6.6^\circ]$  & $[4.5^\circ,\,12.6^\circ]$ & $[3.9^\circ,\,12.3^\circ]$ &&
                           $[5.0^\circ,\,14.4^\circ]$ & $[0,\,12.0^\circ]$         & $[0,\,12.3^\circ]$         \\ \hline\hline
  ${\bf B}_1$$^\ast$     & $[1.6^\circ,\,6.6^\circ]$  & $[4.5^\circ,\,12.6^\circ]$ & $[0,\,26.6^\circ]$         &&
                           $[5.0^\circ,\,14.4^\circ]$ & $[0,\,12.0^\circ]$         & $[0,\,23.7^\circ]$         \\ \hline
  ${\bf B}_2$$^\dag$     & $[1.6^\circ,\,6.6^\circ]$  & $[4.5^\circ,\,12.6^\circ]$ & $[4.0^\circ,\,19.1^\circ]$ &&
                           $[5.0^\circ,\,14.4^\circ]$ & $[0.7^\circ,\,12.0^\circ]$ & $>0.6^\circ$               \\ \hline
  ${\bf B}_3$$^\ast$     & $[1.6^\circ,\,6.6^\circ]$  & $[0,\,26.6^\circ]$         & $[3.9^\circ,\,12.3^\circ]$ &&
                           $[5.0^\circ,\,14.4^\circ]$ & $[0,\,23.0^\circ]$         & $[0,\,12.3^\circ]$         \\ \hline
  ${\bf B}_4$$^\dag$     & $[1.6^\circ,\,6.6^\circ]$  & $[4.2^\circ,\,21.6^\circ]$ & $[3.9^\circ,\,12.3^\circ]$ &&
                           $[5.0^\circ,\,14.4^\circ]$ & $>0.6^\circ$               & $[0.6^\circ,\,12.3^\circ]$ \\ \hline
  ${\bf B}_5$$^\ddag$    & $[0,\,7.5^\circ]$          & $[4.5^\circ,\,12.6^\circ]$ & $[3.9^\circ,\,12.3^\circ]$ &&
                           $-$                        & $[1.1^\circ,\,12.0^\circ]$ & $[0.6^\circ,\,12.3^\circ]$ \\ \hline
  ${\bf B}_6$$^\ddag$    & $[0,\,8.2^\circ]$          & $[4.5^\circ,\,12.6^\circ]$ & $[3.9^\circ,\,12.3^\circ]$ &&
                           $-$                        & $[0.5^\circ,\,12.0^\circ]$ & $[1.3^\circ,\,12.3^\circ]$ \\ \hline\hline
  ${\bf C}_1$$^\S$       & $[1.6^\circ,\,6.6^\circ]$  & $[4.5^\circ,\,12.6^\circ]$ & $-$                        &&
                           $[5.0^\circ,\,14.4^\circ]$ & $[0,\,12.0^\circ]$         & $-$                        \\ \hline
  ${\bf C}_2$$^\S$       & $[1.6^\circ,\,6.6^\circ]$  & $-$                        & $[3.9^\circ,\,12.3^\circ]$ &&
                           $[5.0^\circ,\,14.4^\circ]$ & $-$                        & $[0,\,12.3^\circ]$         \\ \hline
  ${\bf C}_3$            & $-$                        & $[4.5^\circ,\,12.6^\circ]$ & $[3.9^\circ,\,12.3^\circ]$ &&
                           $-$                        & $[0,\,12.0^\circ]$         & $[0,\,12.3^\circ]$         \\ \hline\hline
  ${\bf D}_1$$^\P$       & $[1.6^\circ,\,6.6^\circ]$  & $[0.7^\circ,\,26.6^\circ]$ & $>2.4^\circ$               &&
                           $[5.0^\circ,\,14.4^\circ]$ & $[0.7^\circ,\,23.0^\circ]$ & $>0.2^\circ$               \\ \hline
  ${\bf D}_2$$^\P$       & $[1.6^\circ,\,6.6^\circ]$  & $>2.5^\circ$               & $[0.7^\circ,\,26.6^\circ]$ &&
                           $[5.0^\circ,\,14.4^\circ]$ & $>0.2^\circ$               & $[0.7^\circ,\,23.7^\circ]$ \\ \hline
  ${\bf D}_3$$^\flat$    & $[0.3^\circ,\,7.5^\circ]$  & $[4.5^\circ,\,12.6^\circ]$ & $-$                        &&
                           $-$                        & $[1.1^\circ,\,12.0^\circ]$ & $-$                        \\ \hline
  ${\bf D}_4$$^\natural$ & $[0,\,8.2^\circ]$          & $[4.5^\circ,\,12.6^\circ]$ & $[4.0^\circ,\,19.1^\circ]$ &&
                           $-$                        & $[0.5^\circ,\,12.0^\circ]$ & $>0.6^\circ$               \\ \hline
  ${\bf D}_5$$^\flat$    & $[0.3^\circ,\,8.2^\circ]$  & $-$                        & $[3.9^\circ,\,12.3^\circ]$ &&
                           $-$                        & $-$                        & $[1.3^\circ,\,12.3^\circ]$ \\ \hline
  ${\bf D}_6$$^\natural$ & $[0,\,7.8^\circ]$          & $[4.2^\circ,\,21.6^\circ]$ & $[3.9^\circ,\,12.3^\circ]$ &&
                           $-$                        & $>0.6^\circ$               & $[0.7^\circ,\,12.3^\circ]$ \\ \hline\hline
  ${\bf E}_1$            & $[1.6^\circ,\,6.6^\circ]$  & $[0,\,26.6^\circ]$         & $[0,\,26.6^\circ]$         &&
                           $[5.0^\circ,\,14.4^\circ]$ & $[0,\,23.0^\circ]$         & $[0,\,23.7^\circ]$         \\ \hline
  ${\bf E}_2$$^\sharp$   & $[0,\,7.5^\circ]$          & $[4.5^\circ,\,12.6^\circ]$ & $[4.0^\circ,\,19.1^\circ]$ &&
                           $-$                        & $[1.1^\circ,\,12.0^\circ]$ & $>0.6^\circ$               \\ \hline
  ${\bf E}_3$$^\sharp$   & $[0,\,8.2^\circ]$          & $[4.2^\circ,\,21.6^\circ]$ & $[3.9^\circ,\,12.3^\circ]$ &&
                           $-$                        & $>0.6^\circ$               & $[1.2^\circ,\,12.3^\circ]$ \\ \hline\hline
  ${\bf F}$              & $[0.3^\circ,\,4.4^\circ]$  & $>0.8^\circ$               & $>0.7^\circ$               &&
                           $[0,\,70.8^\circ]$         & $>0.9^\circ$               & $>1.4^\circ$               \\ \hline\hline
  \end{tabular}
  \end{table}

  \item
  Class ${\bf C}$

  We first consider the texture ${\bf C_1}$, in which $m_{e\mu}=0$ in addition to $m_{ee}=m_{\mu\mu}=0$. The three corresponding Eqs.~(\ref{eqn:a}), (\ref{eqn:b}), and (\ref{eqn:d}) are irrelevant to the ASM angle $\theta_{34}$; actually even the three exact vanishing conditions $m_{ee,\,\mu\mu,\,\tau\tau}=0$ without any approximation does not involve $\theta_{34}$ and thus $\theta_{34}$ are completely unconstrained. The predicted ranges of $\theta_{14,\,24}$ are exactly the same as in the texture ${\bf A}$, assuming $m_0=0.01$~eV,
  \begin{equation}
  \begin{array}{lcclc}
  1.6^\circ < \theta_{14} < 6.6^\circ & (\mbox{NH}) & \quad\mbox{ or }\quad &
  5.0^\circ < \theta_{14} < 14.4^\circ  & (\mbox{IH}) \,, \\
  4.5^\circ < \theta_{24} < 12.6^\circ & (\mbox{NH}) & \quad\mbox{ or }\quad &
  0 < \theta_{24} < 12.0^\circ  & (\mbox{IH}) \,,
  \end{array}
  \end{equation}
  whereas Eq.~(\ref{eqn:d}) implies
  \begin{equation}
  \begin{array}{lcclc}
  0 < \sqrt{\theta_{14}\theta_{24}} \lesssim 7.6^\circ & (\mbox{NH}) & \quad\mbox{ or }\quad &
  0 < \sqrt{\theta_{14}\theta_{24}} \lesssim 12.8^\circ  & (\mbox{IH}) \,.
  \end{array}
  \end{equation}
  Obviously, these three vanishing conditions are consistent with each other. When the neutrino spectrum is inverted, the vanishing nondiagonal element $m_{e\mu}$ is largely incompatible with the two diagonal conditions: When $\alpha\sim\pi$ Eq.~(\ref{eqn:d2}) produces larger values of $\theta_{14,\,24}$ than Eqs.~(\ref{eqn:a}) and (\ref{eqn:b}), whereas when $\alpha\sim0$, it leads to smaller values of $\theta_{14,\,24}$. Thus IH is highly disfavored by the texture ${\bf C_1}$. For the NH spectrum, however, the three vanishing conditions are compatible and the phase $\alpha\sim\pi$ is favored due to the same reason as stated for the textures {\bf A} and ${\bf B_1}$.

  At LO, the texture ${\bf C_2}$ does not involve the angle $\theta_{24}$ and ${\bf C_3}$ is irrelevant to $\theta_{14}$.\footnote{At next-to-leading (NLO) order, the two textures ${\bf C_{2,\,3}}$ involve all the three ASM angles, but the formulas are rather complex. Thus we do not show the formulas of $s_{24}$ of ${\bf C_2}$ and $s_{14}$ of ${\bf C_1}$ in Table~\ref{table:asm-exp}. The preferred ranges of the two angles can be easily recognized in the numerical calculation below.} The expressions and numerical ranges of the other ASM angles for ${\bf C_2}$ and ${\bf C_3}$ are given, respectively, in Tables~\ref{table:asm-exp} and \ref{table:asm-num}. ${\bf C_2}$ is related to ${\bf C_1}$ by the $\mu-\tau$ symmetry, and it favors the NH spectrum and the values of $\alpha$ around $\pi$. For texture ${\bf C_3}$, as a result of the different signs of the cross terms in Eq.~(\ref{eqn:f}) from that in (\ref{eqn:b}) and (\ref{eqn:c}), no matter $\beta$ is $\sim0$ or $\sim\pi$, it is challenging for (\ref{eqn:f}) to keep consistent with (\ref{eqn:b}) and (\ref{eqn:c}) and then this texture is highly disfavored.

  \item
  Class ${\bf D}$

  In the first texture ${\bf D_1}$ of this class, $m_{ee}=0$ and $m_{e\mu}=m_{\mu\tau}=0$. We can easily obtain the analytical formulas of the three ASM angles from Eqs.~(\ref{eqn:s14}), (\ref{eqn:s1424}), and (\ref{eqn:s2434}),
  \begin{subequations}
  \begin{align}
  s_{14}  &= a \,, \label{eqn:D1s14} \\
  s_{24}  &= d^2/a \,, \label{eqn:D1s24} \\
  s_{34}  &= af^2/d^2 \,.
  \end{align}
  \end{subequations}
  Assuming $m_0=0.01$ eV, the numerical ranges of th ASM angles are,
  \begin{equation}
  \label{eqn:D1}
  \begin{array}{lcclc}
  1.6^\circ < \theta_{14} < 6.6^\circ & (\mbox{NH}) & \quad\mbox{ or }\quad &
  5.0^\circ < \theta_{14} < 14.4^\circ  & (\mbox{IH}) \,, \\
  0.7^\circ < \theta_{24} < 26.6^\circ & (\mbox{NH}) & \quad\mbox{ or }\quad &
  0.7^\circ < \theta_{24} < 23.0^\circ  & (\mbox{IH}) \,, \\
  2.4^\circ < \theta_{34} & (\mbox{NH}) & \quad\mbox{ or }\quad &
  0.2^\circ < \theta_{34} & (\mbox{IH}) \,.
  \end{array}
  \end{equation}
  Originally, the lower bounds on $\theta_{24}$ for both NH and IH spectrums are vanishing. However, as the constraint on $\theta_{34}$ is obtained by means of $s_{34} \propto 1/s_{24}$, to ensure that $s_{34} \leq 1$, a cutoff is imposed on the lower end of $s_{24}$, which is estimated to be $0.7^\circ$ as shown in Eq.~(\ref{eqn:D1}). Due to the opposite sign of the cross term $m_1m_2$ in Eqs.~(\ref{eqn:a}) and (\ref{eqn:d2}), the IH spectrum is highly disfavored by the texture ${\bf D_1}$; for NH the phase $\alpha\sim\pi$ is much more favored than $\alpha\sim0$.

  For the textures ${\bf D_2}-{\bf D_6}$, see Tables~\ref{table:asm-exp} and \ref{table:asm-num} for the expressions and expected numerical ranges of the ASM angles, which reveal the relation of ${\bf D_1}-{\bf D_2}$, ${\bf D_3}-{\bf D_5}$, and ${\bf D_4}-{\bf D_6}$ under the approximate $\mu-\tau$ symmetry as we expect. With the method of rough estimation frequently applied above, we can foresee that only the texture ${\bf D_2}$ disfavor IH. In addition, the textures ${\bf D_{2,\,4,\,6}}$ require that $\alpha\sim\pi$, while ${\bf D_{3,\,5}}$ are more acceptable if $\alpha\sim0$ and $\beta\sim\pi$.

  \item
  Class ${\bf E}$

  In the texture ${\bf E_1}$, $m_{ee}=0$ and $m_{e\mu}=m_{e\tau}=0$. After applying Eq.~(\ref{eqn:s14}), $s_{24}$ and $s_{34}$ can be written as functions of $s_{14}$ and then
  \begin{subequations}
  \begin{align}
  s_{14}   &= a \,, \label{eqn:E1s14} \\
  s_{24}   &= d^2/a \,, \label{eqn:E1s24} \\
  s_{34}   &= e^2/a \,.
  \end{align}
  \end{subequations}
  With the lightest neutrino mass $m_0=0.01$ eV, estimated ranges of the ASM angles are, for both NH and IH spectrums,
  \begin{equation}
  \begin{array}{lcclc}
  1.6^\circ < \theta_{14} < 6.6^\circ & (\mbox{NH}) & \quad\mbox{ or }\quad &
  5.0^\circ < \theta_{14} < 14.4^\circ  & (\mbox{IH}) \,, \\
  0 < \theta_{24} < 26.6^\circ & (\mbox{NH}) & \quad\mbox{ or }\quad &
  0 < \theta_{24} < 23.0^\circ  & (\mbox{IH}) \,, \\
  0 < \theta_{34} <26.6^\circ  & (\mbox{NH}) & \quad\mbox{ or }\quad &
  0 < \theta_{34} < 23.7^\circ & (\mbox{IH}) \,.
  \end{array}
  \end{equation}
  Based on the argument analogical to that for the texture ${\bf D_1}$, IH is high suppressed compared to NH, in which the preferred value of $\alpha$ is around $\pi$.

  The expressions and numerical ranges of the ASM angles in the textures ${\bf E_{2,\,3}}$ are presented, respectively, in Tables~\ref{table:asm-exp} and \ref{table:asm-num}. Obviously, the approximate $\mu-\tau$ symmetry holds for the two textures, and both the two neutrino mass hierarchies are compatible with the two structures. For the NH spectrum, to obtain small ASM angles, $\alpha\sim0$ and $\beta\sim\pi$ is required in both the two textures, while for IH, $\alpha$ tends to be around $\pi$.

  \item
  Texture ${\bf F}$

  In this texture the three nondiagonal elements in the active sector of $M_\nu$ vanish, i.e., $m_{e\mu}=m_{e\tau}=m_{\mu\tau}=0$. Based on Eqs.~(\ref{eqn:s1424})---(\ref{eqn:s2434}), the three ASM angles can be expressed as
  \begin{eqnarray}
  s_{14} &=& de/f \,, \nonumber \\
  s_{24} &=& df/e \,, \nonumber \\
  s_{34} &=& ef/d \,,
  \end{eqnarray}
  Assuming as above $m_0=0.01$ eV, we can estimate the ASM angles,
  \begin{equation}
  \begin{array}{lcclc}
  0.3^\circ < \theta_{14} < 4.4^\circ & (\mbox{NH}) & \quad\mbox{ or }\quad &
  0.1^\circ < \theta_{14} < 70.8^\circ  & (\mbox{IH}) \,, \\
  0.8^\circ < \theta_{24} & (\mbox{NH}) & \quad\mbox{ or }\quad &
  0.9^\circ < \theta_{24}  & (\mbox{IH}) \,, \\
  0.7^\circ < \theta_{34}  & (\mbox{NH}) & \quad\mbox{ or }\quad &
  1.0^\circ < \theta_{34}  & (\mbox{IH}) \,.
  \end{array}
  \end{equation}
  As seen in Eqs.~(\ref{eqn:d})-(\ref{eqn:f}), all the three CP violating phases $\alpha$, $\beta$, and $\delta_{13}$ play an important role in pinning down the neutrino spectrum. It is not so transparent as the textures above to figure out the preferred neutrino spectrum and the favored regions of the phases. More phenomenological discussions of this textures are postponed to the next section, wherein the numerical analysis can help us to further investigate this texture.
\end{itemize}

After the detailed investigation of the neutrino spectrum, ASM angles and CP violating phases of the 20 type II textures, we now estimate both the effective neutrino masses in both tritium beta decay and $0\nu\beta\beta$. With $m_0\sim0.01$~eV, the contribution from the sterile state dominates the beta decay electron neutrino mass
\begin{equation}
\label{eqn:mbeta}
m_\beta = \sqrt{ m_1^2c^2_{12}c^2_{13}c^2_{14} +m_2^2s^2_{12}c^2_{13}c^2_{14}  +m_4^2s^2_{14} } \sim 0.1\;{\rm eV} \,.
\end{equation}
In Eq.~(\ref{eqn:mbeta}) we neglect the contribution from $m_3$, as its is highly suppressed by the smallness of reactor mixing $s_{13}^2$. Assuming explicitly $m_0=0.01$~eV as well as the $3\sigma$ ranges of the active neutrino parameters and $\Delta m^2_{41}$ and $\theta_{14}$ listed in Tables~\ref{table:nu-fit} and \ref{table:sn-fit}, one can more accurately predict the mass $m_\beta$ in general (3+1) neutrino models:
\begin{equation}
\label{eqn:mbeta-num}
  \begin{array}{lcclc}
  91~{\rm meV} < m_\beta < 420~{\rm meV} & (\mbox{NH}) & \quad\mbox{ or }\quad &
  102~{\rm meV} < m_\beta < 423~{\rm meV} & (\mbox{NH}) \,.
  \end{array}
\end{equation}
With the help of Eqs~(\ref{eqn:142434}) and (\ref{eqn:abcdef}), we calculate the effective mass $m_\beta$ in all the 20 textures, as collected in Table~\ref{table:mbeta-mee}. For both NH and IH spectrums, $m_\beta$ in all the textures are of order 0.1~eV, consistent with the rough estimation. Furthermore, the predictions presented in the table overlap partially with Eq.~(\ref{eqn:mbeta-num}), which implies again that these type II textures are compatible with experiments. The lower limits $\sim\!13$~meV and $\sim\!48$~meV correspond to a vanishing $\theta_{14}$ and denote the minimum of the contribution from the active neutrinos with the NH and IH spectrums, respectively. If the neutrino spectrum is inverted, $m_\beta$ is mildly augmented as a result of heavier $m_{1,\,2}$. The ASM angle $\theta_{14}$ in the texture ${\bf C_3}$ are not restricted at LO, thus the neutrino mass $m_\beta$ for ${\bf C_3}$ are left blank in Table~\ref{table:mbeta-mee}. The designed discovery sensitivity of the KATRIN experiment is about 0.35~eV~\cite{KATRIN} and it seems possible for KATRIN to distinguish the neutrino mass hierarchies in these textures, if a positive signal is observed in this experiment.
\begin{table}[tp]
  \centering
  \caption{LO estimations of the effective electron neutrino mass $m_\beta$ in beta decay (left table) and the effective Majorana electron neutrino mass $\langle m_{ee} \rangle$ in $0\nu\beta\beta$ (right table), with all in units of meV. See text for details.}
  \label{table:mbeta-mee}
  \hspace{.1cm}
  \begin{tabular}{ccc}
  \hline\hline
  Textures & NH & IH \\ \hline
  ${\bf A,\, B_{1,\,2,\,3,\,4},\,  C_{1,\,2},\, D_{1,\,2},\, E_1 }$ & $36-135$ &  $112-294$  \\ \hline
  ${\bf B_{5},\, D_{3},\, E_2 }$ & $13-152$ & $\gtrsim48$ \\ \hline
  ${\bf B_{6},\, D_{5},\, E_3 }$ & $13-166$ & $\gtrsim48$ \\ \hline
  ${\bf D_{4} }$ &$13-166$ & $\gtrsim48$ \\ \hline
  ${\bf D_{6} }$ & $13-159$ & $\gtrsim48$ \\ \hline
  ${\bf F }$ & $13-90$ & $\gtrsim48$ \\ \hline
  ${\bf C_{3} }$ & $-$ & $-$ \\ \hline\hline
  \end{tabular} \hspace{.3cm}
  \begin{tabular}{ccc}
  \hline\hline
  Textures & NH & IH \\ \hline
  ${\bf A,\, B_{1,\,2,\,3,\,4},\,  C_{1,\,2},\, D_{1,\,2},\, E_1 }$ & $0$ &  $0$  \\ \hline
  ${\bf B_{5},\, D_{3},\, E_2 }$ & $0-15$ & $\gtrsim47$ \\ \hline
  ${\bf B_{6},\, D_{5},\, E_3 }$ & $0-16$ & $\gtrsim47$ \\ \hline
  ${\bf D_{4} }$ &$0-22$ & $\gtrsim47$ \\ \hline
  ${\bf D_{6} }$ & $0-21$ & $\gtrsim47$ \\ \hline
  ${\bf F }$ & $0-13$ & $\gtrsim47$ \\ \hline
  ${\bf C_{3} }$ & $-$ & $-$ \\ \hline\hline
  \end{tabular}
  \hspace{.1cm}
\end{table}

With regard to the effective Majorana neutrino mass $\langle m_{ee} \rangle$ in $0\nu\beta\beta$, if $m_0=0.01$~eV, neglecting the highly suppressed $m_3s^2_{13}$ term,
\begin{equation}
\langle m_{ee} \rangle = m_{ee} = \left| m_1c^2_{12}c^2_{13}c^2_{14} +m_2s^2_{12}c^2_{13}c^2_{14}e^{i\alpha} +m_4s^2_{14}e^{i(-2\delta_{14}+\gamma)} \right|  \,.
\end{equation}
Rough estimation of the order of magnitude tells us that $\langle m_{ee} \rangle$ is of order $\mathcal{O}(10\;{\rm meV})$, which is just of the same order as the sensitivity of the next generation $0\nu\beta\beta$ experiments~\cite{0nubetabeta-review}. The leading order predicted ranges of $\langle m_{ee} \rangle$ for the 20 textures are listed in Table~\ref{table:mbeta-mee}. The ten textures ${\bf A,\, B_{1,\,2,\,3,\,4},\, C_{1,\,2},\, D_{1,\,2},\, E_1 }$ assume $ m_{ee}=0 $, thus definitely they cannot produce any signal in the $0\nu\beta\beta$ experiments. With help of the Majorana phases $\alpha$ and $\gamma$, the nine textures ${\bf B_{5,\,6},\, D_{3,\,4,\,5,\,6},\, E_{2,\,3},\, F }$ with NH spectrum can lead to vanishing Majorana neutrino mass, and the maximum value of $\langle m_{ee} \rangle$ in these textures at the leading order is about (10--20)~meV. For IH, $m_{1,\,2}$ become heavier and it is impossible for the states $m_{1,\,2,\,4}$ to completely cancel out each other, thus we arrive at a nonvanishing lower bound of $\sim47$~meV in the nine textures. At LO the texture ${\bf C_3}$ cannot predict $\langle m_{ee} \rangle$, however, if it is compatible with the SBL experiments, the prediction of ${\bf C_3}$ on $\langle m_{ee} \rangle$ is about the same as the nine textures such as ${\bf B_5}$. The most meaningful phenomenological features in Table~\ref{table:mbeta-mee} is that with $m_0=0.01$~eV the NH spectrums of the ten textures ${\bf B_{5,\,6},\, C_3,\, D_{3,\,4,\,5,\,6},\, E_{2,\,3},\, F }$ are clearly distinguishable from the IH spectrums, and it is promising to test the neutrino mass hierarchies in these textures by means of the next generation $0\nu\beta\beta$ experiments~\cite{0nubetabeta-review}. The phenomenological implications of these textures on the single and double beta decays will be further investigated in the following section on numerical analysis.

\section{Numerical analysis}
\label{sec:calculation}

We now turn to numerical analysis of the 120 textures of $M_\nu$ with three zeros in the (3+1) neutrino model. In this section we first state in detail how the numerical calculation is implemented and then analyze systematically the numerical results of the type I and II textures. A number of scatter plots are depicted to illustrate the phenomenological consequences of these textures: neutrino mass spectrum, ASM angles, CP violating angles, and the nonoscillation quantities, i.e., sum of neutrino mass $\Sigma_m$, effective neutrino mass $m_\beta$ in beta decay and the Majorana neutrino mass $\langle m_{ee} \rangle$ in $0\nu\beta\beta$.

\subsection{Calculation procedure}

The numerical calculation is implemented as follows:
\begin{itemize}
  \item
  The three known active neutrino mixing angles $\theta_{12,\,23,\,13}$ are allowed to vary randomly in their $3\sigma$ ranges presented in Table~\ref{table:nu-fit}; the three ASM angles $\theta_{14,\,24,\,34}$ and three Dirac-type CP violating phases $\delta_{13,\,14,\,24}$ are viewed as ``unknown'' parameters and they vary randomly in the range [0, $\pi/2$] and [0, $2\pi$], respectively.
  \item
  For each randomly generated group of the nine physical parameters given above, we calculate the corresponding $\lambda_{21,\,31,\,41}$ and the two resultant neutrino mass square difference ratios $R_{\nu1}$ and $R_{\nu2}$, based on Eqs.~(\ref{matrix:U}), (\ref{lambdaNN}), (\ref{lambdaNNN}), and (\ref{Rnu}). At this stage, the three neutrino mass ratios $m_2/m_1$, $m_3/m_1$, $m_4/m_1$ and three Majorana-type phases $\alpha$, $\beta$, $\gamma$ can be obtained according to, respectively, Eqs.~(\ref{massratio}) and (\ref{Majoranaphase}). If $R_{\nu1}$ and $R_{\nu2}$ fall simultaneously inside the ranges determined by the data on mass square differences in Tables~\ref{table:nu-fit} and \ref{table:sn-fit}
  \begin{eqnarray}
  &&0.0260 < R_{\nu1} < 0.0371 \nonumber \,, \\
  &&2.81\times10^{-5} < R_{\nu2} < 1.14\times10^{-4} \,,
  \end{eqnarray}
  then the corresponding input groups of neutrino parameters are \emph{possibly} compatible with experimental constraints, otherwise, they are excluded by experiments at $3\sigma$ C.L.
  \item
  For the groups of neutrino parameters consistent with experiments obtained above, we randomly generate $\Delta m^2_{21}$ in its $3\sigma$ range. With the mass ratio $m_2/m_1$ obtained above, we can easily get the corresponding absolute neutrino masses $m_{1,\,2}$, and further produce $m_{3,\,4}$ when the mass ratios $m_3/m_1$ and $m_4/m_1$ are also involved. Given the four neutrino masses $m_{1,\,2,\,3,\,4}$, it is \emph{compulsory} to check if the corresponding $|\Delta m^2_{31}|$ and $\Delta m^2_{41}$ do fall into their corresponding $3\sigma$ ranges and discard those that do not.
  \item
  Finally, we obtain all the 16 neutrino physical parameters, among which six are constrained by experimental data: $\theta_{12,\,23,\,13}$, $\Delta m^2_{21}$, $|\Delta m^2_{31}|$, and $\Delta m^2_{41}$, six are randomly generated unconstrained: $\theta_{14,\,24,\,34}$ and $\delta_{13,\,14,\,24}$, and the last four, i.e., $\alpha$, $\beta$, $\gamma$, and $m_0$, are obtained based on the correlation equations implied by the three vanishing elements of $M_\nu$.
\end{itemize}

As aforementioned, we own 10 ``unknown'' parameters and six constraints, it seems more smart and economical if we have four, instead of six, unconstrained parameters in our calculation. However, it is unclear how to choose four from the ten ``unknown'', and more important, obtaining the six ``unknown'' parameters from the correlation equations (\ref{massratio}) and (\ref{Majoranaphase}) would become a troublesome task. Comparatively, our numerical calculation is more straightforward and feasible. Our calculation can be regarded as a method in which two of the six correlation equations are solved with the Monte-Carlo method~\cite{pdg2012}.

As seen in Table~\ref{table:nu-fit}, the $3\sigma$ range of $\theta_{23}$ for NH spectrum is slightly different from the case of IH, and the difference of $3\sigma$ ranges of $|\Delta m^2_{31}|$ for the two mass hierarchies is somewhat even larger. Without bias on any of the two hierarchies, at the beginning of numerical calculation we set the covering ranges for both NH and IH as the $3\sigma$ ranges of $|\Delta m^2_{31}|$ and $\theta_{23}$. Numerically, they are $(2.21 - 2.74) \times 10^{-5}\text{eV}^2$ and $0.36 - 0.68$, respectively. After obtaining the neutrino masses by means of the calculation procedure above, we can immediately identify the mass hierarchy for each group of neutrino parameters. We check whether the mass hierarchy and the values of $|\Delta m^2_{31}|$ and $\theta_{23}$ in a given group of neutrino parameters are allowed by the corresponding experimental data listed in Table~\ref{table:nu-fit}, and discard those that are not.

\subsection{Textures of type I}

As explained in the analytical analysis, the textures with zeros in the sterile sector require large or nearly vanishing ASM angles due to the large mass hierarchy between the active and sterile neutrinos $m_{1,\,2,\,3} \ll m_4$. We numerically calculate all the 100 type I textures and find that all the significant features are consistent with the analytical expectations.

\begin{figure}[pt]
  \centering
  \includegraphics[width=5.6cm]{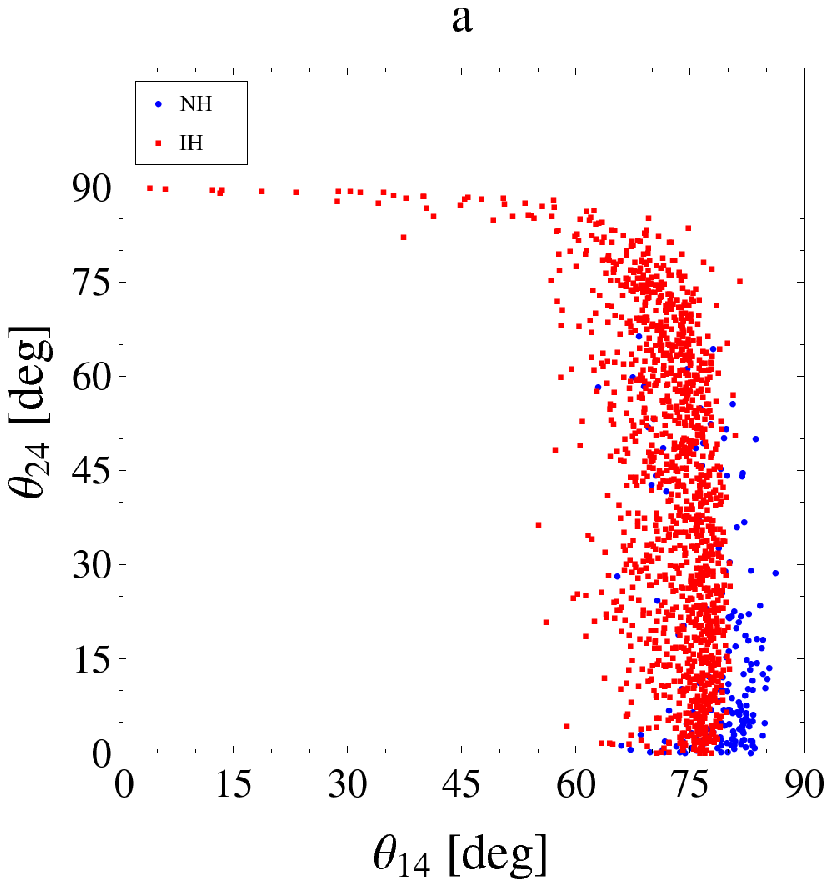}  \hspace{-.5cm}
  \includegraphics[width=5.6cm]{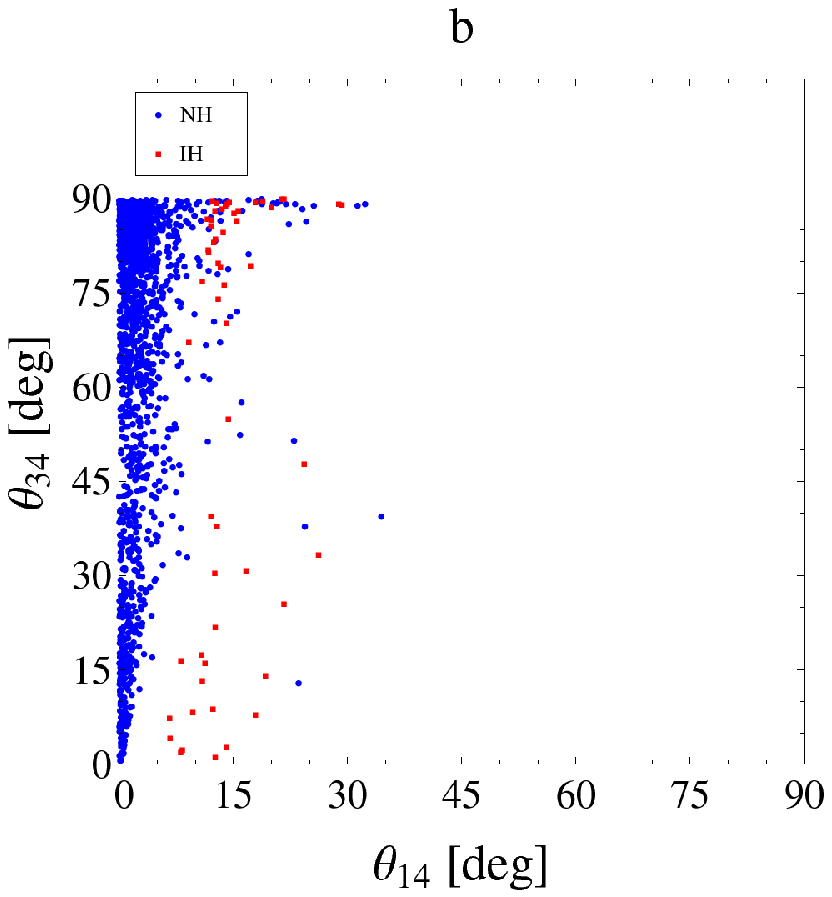} \hspace{-.5cm}
  \includegraphics[width=5.6cm]{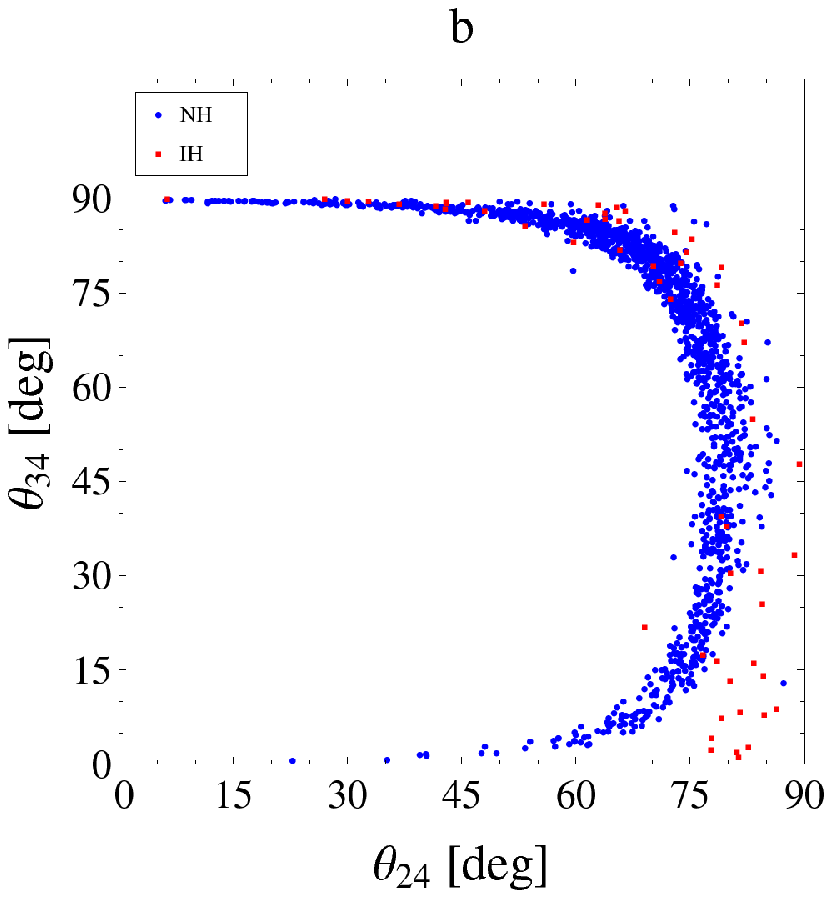}
  \caption{Distributions of the ASM angles for the textures $a$ and $b$ given in Eq.~(\ref{eqn:texture-ab}). The left panel is the $\theta_{14}-\theta_{24}$ distribution in the texture $a$, in units of degree; the middle and right panels show, respectively, the $\theta_{14}-\theta_{34}$ and $\theta_{24}-\theta_{34}$ distributions for the texture $b$. NH and IH are denoted, respectively, as blue filled circles and red squares.}
  \label{fig:type1-1}
\end{figure}

\begin{figure}[pt]
  \centering
  \includegraphics[width=5.6cm]{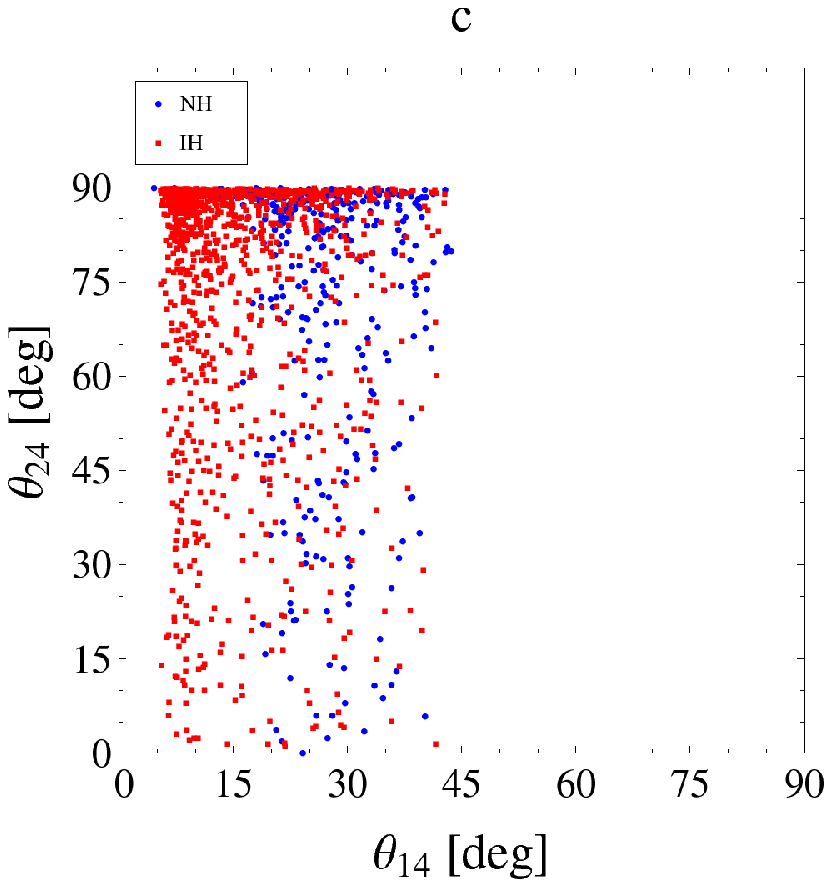} \hspace{-.5cm}
  \includegraphics[width=5.6cm]{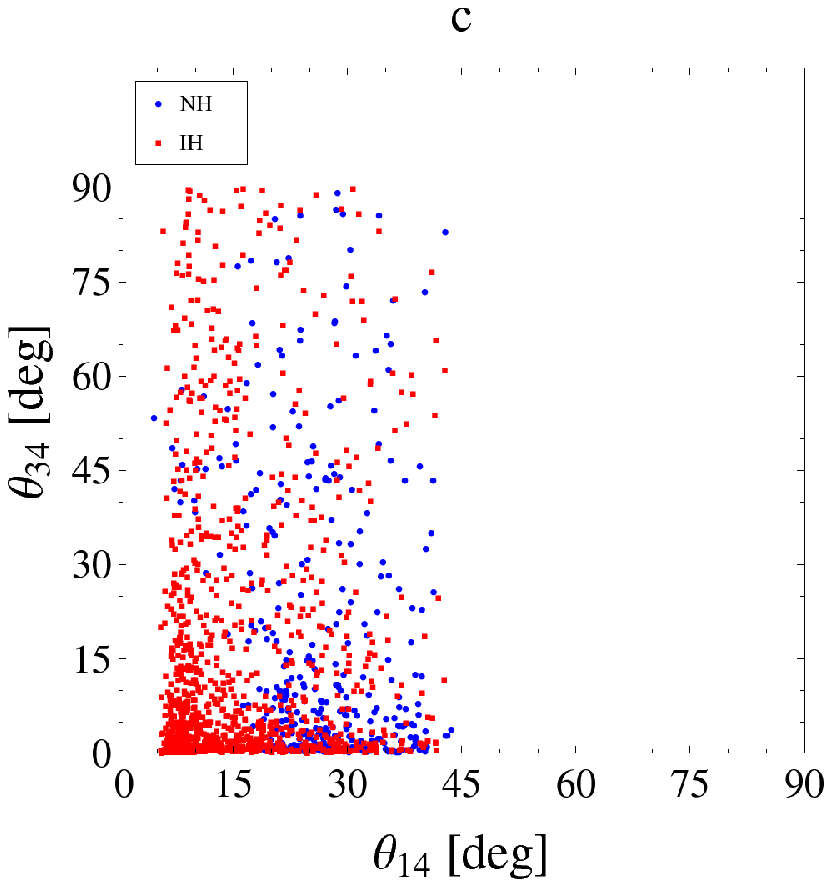}
  \caption{Distributions of the ASM angles for the texture $c$ given in Eq.~(\ref{eqn:texture-c}). The left and right panels are, respectively,  $\theta_{14}-\theta_{24}$ and $\theta_{14}-\theta_{34}$ distributions, in units of degree. NH and IH are denoted, respectively, as blue filled circles and red squares.}
  \label{fig:type1-2}
\end{figure}

As a concrete demonstration, the randomly generated points of the ASM angles of the two textures
\begin{eqnarray}
\label{eqn:texture-ab}
a:\; \left( \begin{array}{cccc}
      \times & \times & \times & \times \\
      \times & \times & \times & 0 \\
      \times & \times & \times & 0 \\
      \times & 0 & 0 & 0 \end{array} \right) \,, \quad
b:\; \left( \begin{array}{cccc}
      0 & \times & \times & 0 \\
      \times & \times & \times & \times \\
      \times & \times & \times & 0 \\
      0 & \times & 0 & \times\end{array} \right)
\end{eqnarray}
are plotted in Fig.~\ref{fig:type1-1}. In the texture $a$ with $m_{ss}=0$, it is transparent that the ASM angle $\theta_{14}$ is very large or, less likely, $\theta_{24}$ is close to the maximum value $\pi/2$, consistent with the analytical expectations~\cite{2zero-sterile}. For the texture $b$, two nondiagonal elements in the sterile sector are zero, i.e., $m_{es,\,\tau s}=0$. We can easily read from the middle and right panels of Fig.~\ref{fig:type1-1} that $\theta_{14}$ is mostly very small and the other two ASM angles are generally very large, which is just predictions of the analytical analysis.
As stated above, the SBL experiments are insensitive to the tauon flavor and thus permit a vanishing $\theta_{34}$. The matrix
\begin{eqnarray}
c:\; \left( \begin{array}{cccc}
 0 & \times & \times & \times \\ \times & \times & 0 & \times \\ \times&0& \times & 0 \\  \times&\times&0& \times
\end{array} \right)
\label{eqn:texture-c}
\end{eqnarray}
has only one zero $m_{\tau s}=0$ in the sterile texture, and the numerical predictions of the three ASM angles are shown in Fig.~\ref{fig:type1-2}. As we expect, $\theta_{34}$ is mostly close to $0$, while $\theta_{24}$ is mostly close to $90^\circ$. By examining the neutrino parameters produced in the numerical calculation, it turns out that only rarely can we obtain the three ASM angles all compatible with SBL experiments, i.e., this texture is highly disfavored, although it has not been completely excluded.

\subsection{Textures of type II}

The analytical analysis reveals that, without compulsorily requiring nearly vanishing or very large ASM angles, it is possible for the type II textures to be compatible with SBL experiments. Now we present the numerical results of the 20 textures, by sort of the various phenomenological implications:
\begin{itemize}
  \item
  The most significant feature which are common to all the 20 textures is that the lightest neutrino mass $m_0$ for both NH and IH is of order $(0.001-0.1)$~eV, as clearly shown in Fig.~\ref{fig:mixing1}. This coincides with the analytical prediction that $m_0$ are required to be around the order 0.01~eV.
  \item
  In Fig.~\ref{fig:mixing1}, we present the numerical predictions of the three ASM angles of texture ${\bf A}$ for both NH and IH spectrums. It is transparent that the ranges of the ASM angles with $m_0=0.01$~eV are almost exactly the same as the analytical predictions. On the whole, if $m_0$ is larger, the ASM angles tend to be larger, in contrast, the angles are smaller for smaller values of $m_0$. The figures of the ASM angles for the textures ${\bf B_1}$, ${\bf C_1}$, ${\bf D_1}$, ${\bf E_1}$, and ${\bf F}$ are shown in Fig.~\ref{fig:mixing2}. It is possible that NLO and higher order corrections beyond Eq.~(\ref{eqn:II}) can widen somewhat the LO allowed ranges of the ASM angles. For instance, in the upper left panel of Fig.~\ref{fig:mixing2}, when $m_0=0.01$~eV, the angle $\theta_{34}$ of ${\bf B_1}$ with NH spectrum can take values larger than the LO upper bound $26.6^\circ$ given in Table~\ref{table:asm-num}. However, the simulations are consistent with the analytical analysis on the whole, which not only verifies the feasibility of the LO analytical analysis, but demonstrate again that the type II textures are more or less compatible with the active and sterile neutrino data.
  \begin{figure}[tp]
  \centering
  \includegraphics[width=6.1cm]{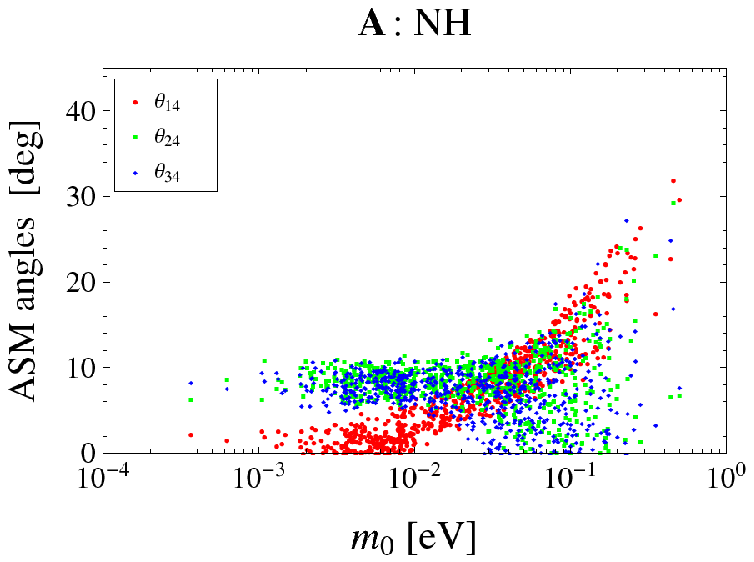} \hspace{.2cm}
  \includegraphics[width=6.1cm]{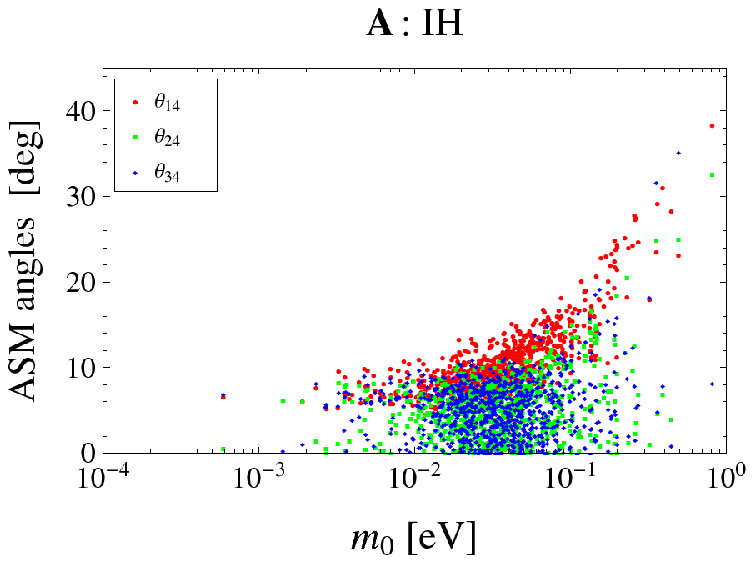}
  \caption{The red filled circles, green squares, and blue diamonds denote, respectively, $\theta_{14}$, $\theta_{24}$, and $\theta_{34}$ in the texture ${\bf A}$ as functions of the lightest neutrino mass $m_0$, with NH (left panel) or IH (right panel) spectrum.}
  \label{fig:mixing1}
  \end{figure}
  \begin{figure}[tp]
  \centering
  \includegraphics[width=6.1cm]{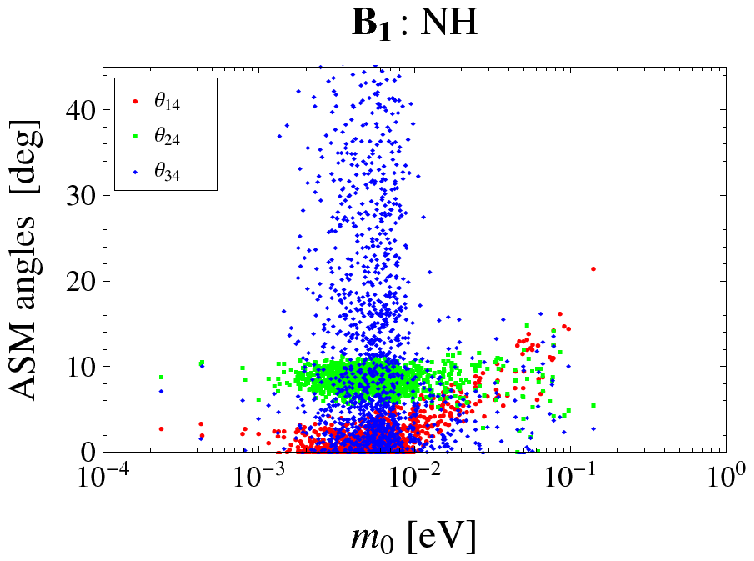} \hspace{.2cm}
  \includegraphics[width=6.1cm]{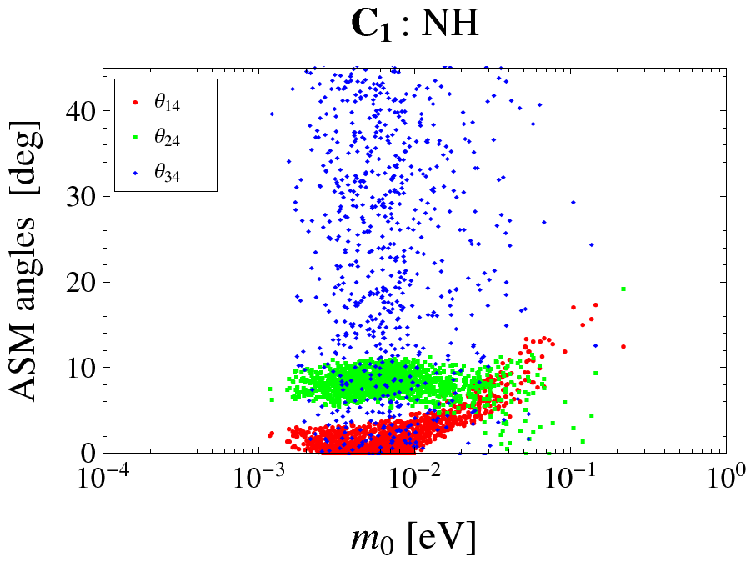} \\ \vspace{.3cm}
  \includegraphics[width=6.1cm]{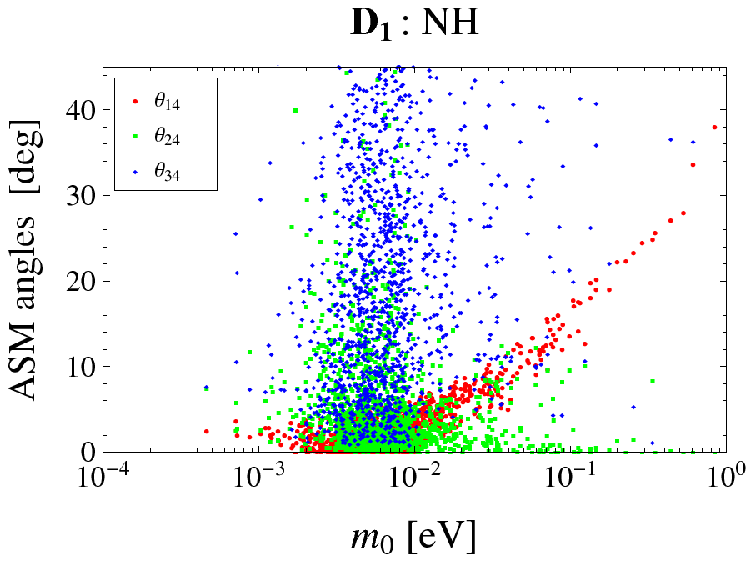} \hspace{.2cm}
  \includegraphics[width=6.1cm]{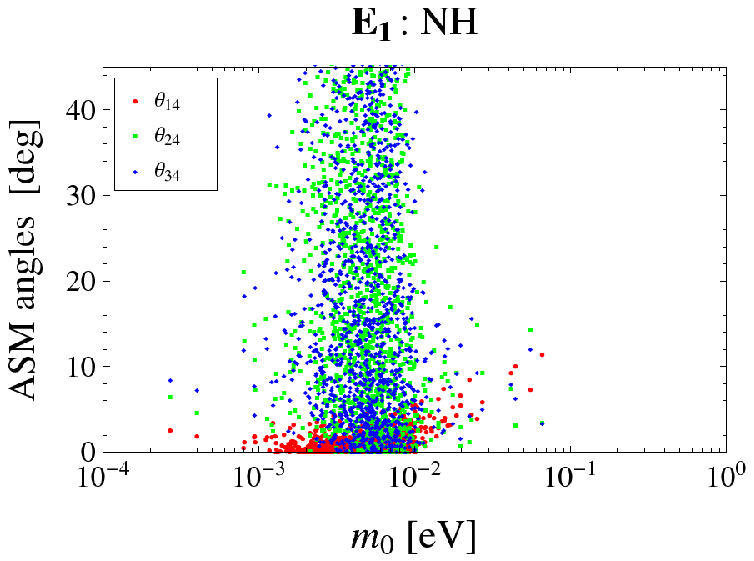} \\ \vspace{.3cm}
  \includegraphics[width=6.1cm]{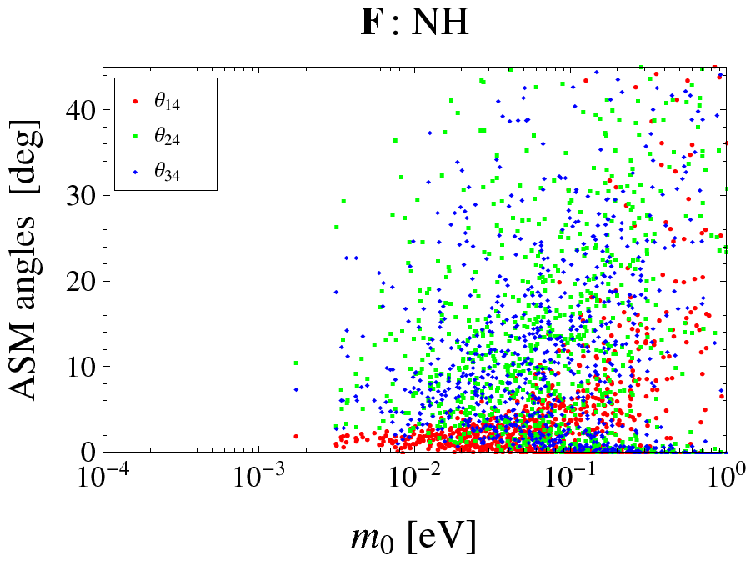} \hspace{.2cm}
  \includegraphics[width=6.1cm]{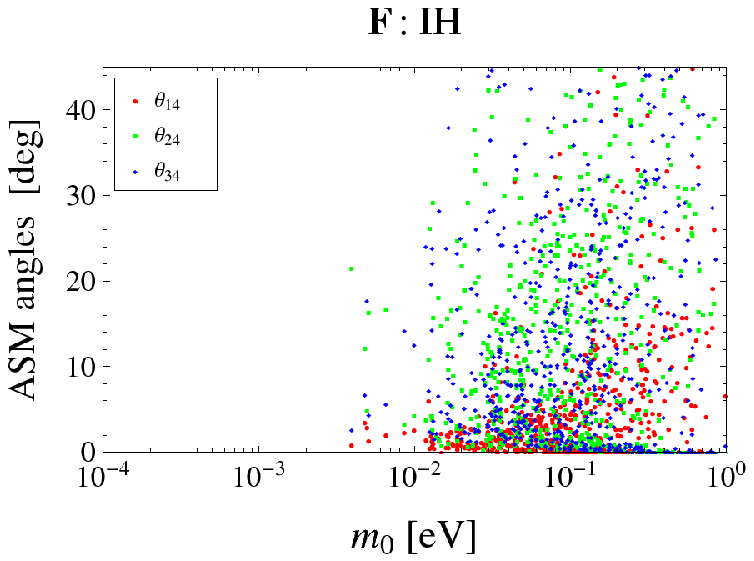} \\ 
  \caption{The red filled circles, green squares, and blue diamonds denote, respectively, $\theta_{14}$, $\theta_{24}$, and $\theta_{34}$ as functions of the lightest neutrino mass $m_0$, in the textures ${\bf B_1}$ (NH, upper left panel), ${\bf C_1}$ (NH, upper right panel), ${\bf D_1}$ (NH, middle left panel), ${\bf E_1}$ (NH, middle right panel), and ${\bf F}$ (lower left panel for NH and low right panel for IH).}
  \label{fig:mixing2}
  \end{figure}

  The three vanishing elements of the texture ${\bf C_1}$ do not involve $\theta_{34}$, thus this angle is completely constrained. In our simulation it distributes evenly in the range $[0,\,\pi/2]$, as we expect, shown in the upper left panel of Fig.~\ref{fig:mixing3}. The other three panels of Fig.~\ref{fig:mixing3} are on the angle $\theta_{24}$ of texture ${\bf C_2}$ (NH) and $\theta_{14}$ of ${\bf C_3}$ (NH and IH) which are restricted at NLO. The numerical calculation reveals that $\theta_{24}$ show a slight preference for larger values around $80^\circ$, whereas $\theta_{14}$ are mostly of the values $70^\circ - 90^\circ$. Consequently, the texture ${\bf C_3}$ is excluded by SBL experiments. In all the discussions below, we will not further consider the texture ${\bf C_3}$.
  \begin{figure}[tp]
  \centering
  \includegraphics[width=6.1cm]{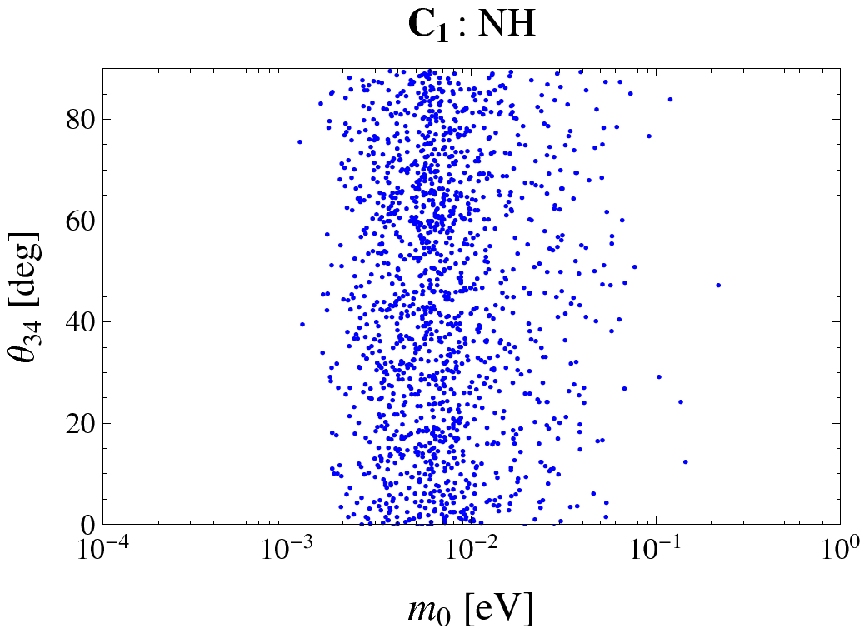} \hspace{.2cm}
  \includegraphics[width=6.1cm]{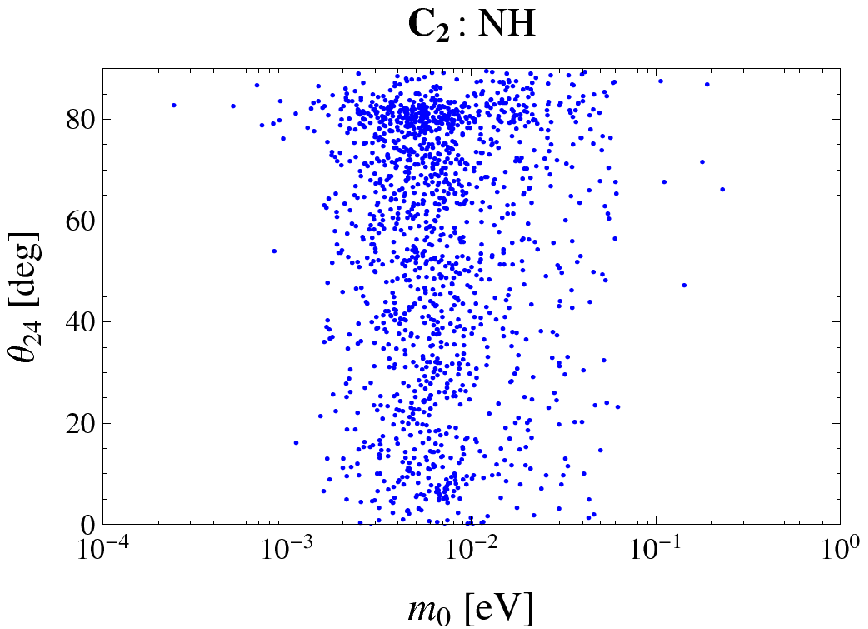} \\ \vspace{.3cm}
  \includegraphics[width=6.1cm]{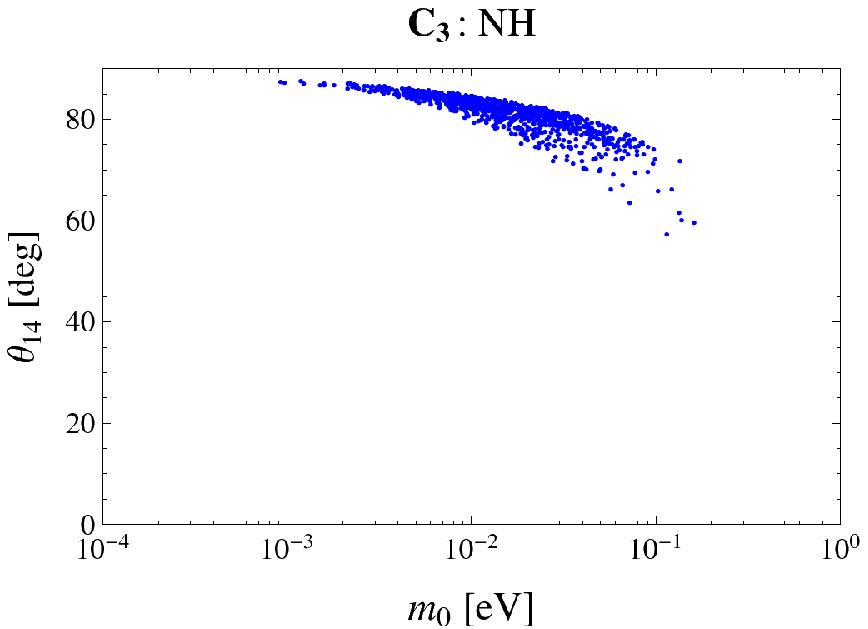} \hspace{.2cm}
  \includegraphics[width=6.1cm]{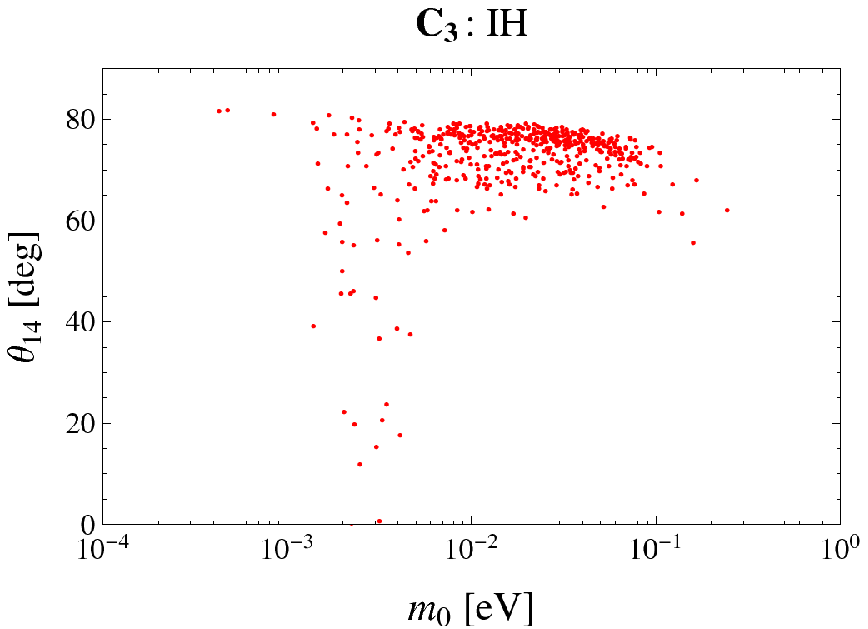} 
  \caption{Distributions of the ASM angle $\theta_{34}$ of the texture ${\bf C_1}$ (NH, upper left panel), $\theta_{24}$ of ${\bf C_2}$ (NH, upper right panel) and $\theta_{14}$ of the texture ${\bf C_2}$ (lower left panel for NH and lower right panel for IH) as functions of the lightest neutrino mass $m_0$. The blue and red points denote, respectively, the NH and IH spectrums.}
  \label{fig:mixing3}
  \end{figure}

  \item
  With regard to the neutrino spectrums of the 19 textures, the numerical calculations are also consistent with the analytical predictions:
  the NH spectrum is much more favored by the seven textures ${\bf B_{1,\,3}}$, ${\bf C_{1,\,2}}$, ${\bf D_{1,\,2}}$, ${\bf E_{1}}$ than IH, whereas both NH and IH are allowed in the other twelve textures ${\bf A}$, ${\bf B_{2,\,4,\,5,\,6}}$, ${\bf D_{3,\,4,\,5,\,6}}$, ${\bf E_{2,\,3}}$, and ${\bf F}$. We deliberately point out the mass spectrum of ${\bf F}$ cannot be clearly determined by analytical analysis.

  \item
  Analytical analysis reveals that some of the CP violating phases are severely constrained, especially the Majorana phases $\alpha$ and $\beta$.
  The preferred ranges of the Dirac and Majorana phases obtained in the numerical calculation are collected in Table~\ref{table:phases}. This table tells us that not only the feasibility of the analytical analysis as well as the predictive power of the type II textures on the phases are verified, but more predictions can be yielded in the numerical calculation on the phases, i.e., the predictions of $\delta_{13,\,24}$ and $\gamma$ presented in the table. In addition to the Dirac phases $\delta_{13}$, the two Majorana phases $\alpha$ and $\beta$ in the texture ${\bf F}$ cannot be clearly predicted by analytical analysis, as we have mentioned above. As representative examples, we present in Fig.~\ref{fig:phases} the numerical results of the phases $\delta_{13}$ in the texture ${\bf F}$ (NH), $\delta_{24}$ in ${\bf B_2}$ (NH), $\alpha$ in ${\bf A}$ (IH) and ${\bf D_5}$ (NH), $\beta$ in ${\bf E_2}$ (NH), and $\gamma$ in ${\bf D_5}$ (NH). All the numerical results are explicitly consistent with the analytical predictions.
  \begin{table}[tp]
  \centering
  \caption{Preferred values of the CP violating phases in the 19 type II textures obtained in the numerical calculation.}
  \label{table:phases}
  \begin{tabular}{lll}
  \hline\hline
  Phase & Preferred values & Textures \\ \hline
  $\delta_{13}$ & $\pi/2$ or $3\pi/2$ & ${\bf F}$  \\ \hline
  $\delta_{24}$ & $0$                 & ${\bf B_{2,\,4}}$, ${\bf D_{4,\,5,\,6}}$, ${\bf E_{2,\,3}}$ and ${\bf F}$, all with NH spectrums  \\ \cline{2-3}
                & $0$ or $\pi$        & ${\bf A}$ (NH); ${\bf F}$ (IH) \\ \cline{2-3}
                & $\pi/2$ or $3\pi/2$ & ${\bf E_1}$  \\ \hline
  $\alpha$      & $0$                 &  ${\bf D_{3,\,5}}$; ${\bf B_{5,\,6}}$ (NH), ${\bf E_{2,\,3}}$ (NH)  \\ \cline{2-3}
                & $\pi$               & ${\bf A}$, ${\bf B_{1,\,2,\,3,\,4}}$, ${\bf C_{1,\,2}}$, ${\bf D_{1,\,2,\,4,\,6}}$, ${\bf E_{1}}$, ${\bf F}$;
                                        ${\bf E_{2,\,3}}$ (IH) \\ \hline
  $\beta$ & $0$ & ${\bf F}$ \\ \cline{2-3}
        & $\pi$                 &  ${\bf D_{3,\,5}}$; ${\bf B_{5,\,6}}$ (NH), ${\bf E_{2,\,3}}$ (NH)  \\ \hline
  $\gamma$ & $\pi$ & ${\bf D_5}$ \\ \hline\hline
  \end{tabular} \hspace{.3cm}
  \end{table}
  \begin{figure}[tp]
  \centering
  \includegraphics[width=6.1cm]{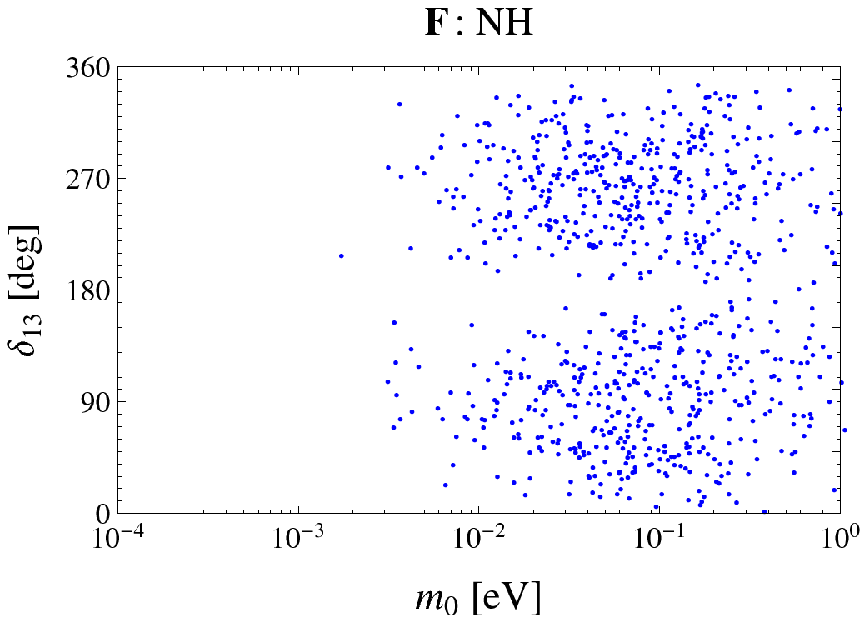} \hspace{.2cm}
  \includegraphics[width=6.1cm]{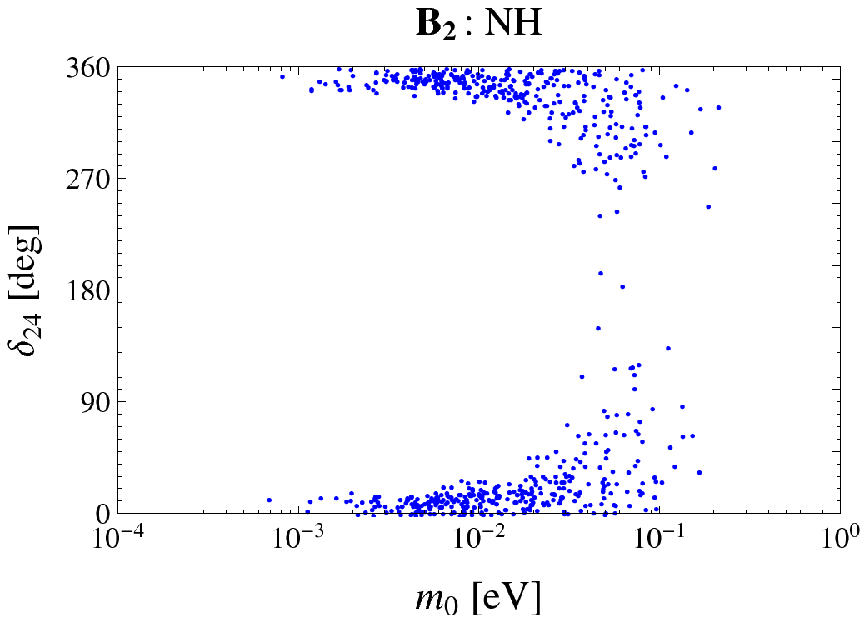} \\ \vspace{.3cm}
  \includegraphics[width=6.1cm]{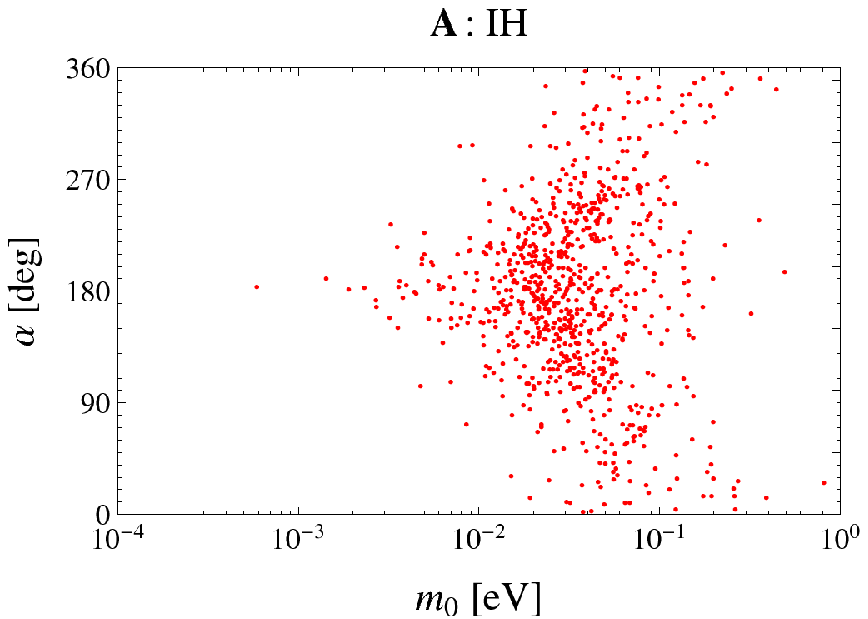} \hspace{.2cm}
  \includegraphics[width=6.1cm]{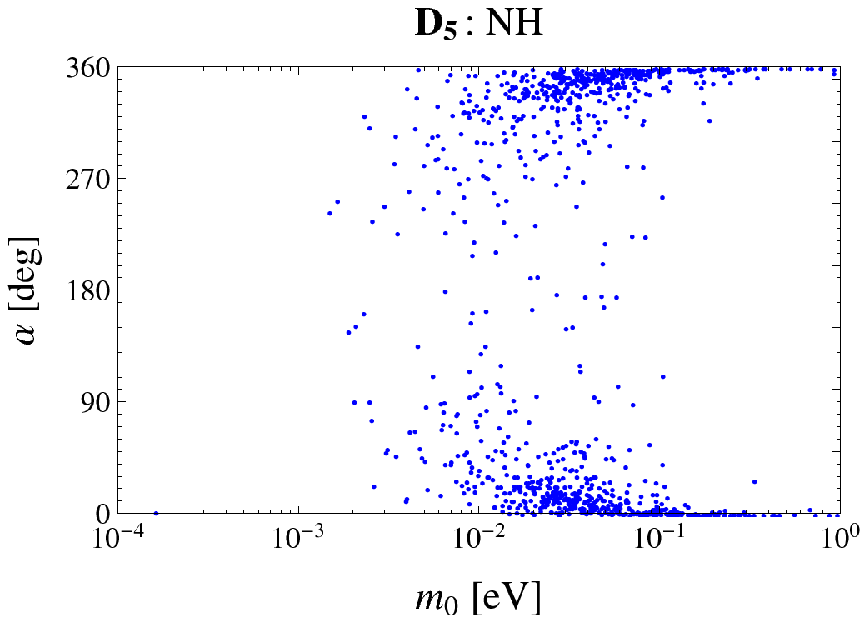} \\ \vspace{.3cm}
  \includegraphics[width=6.1cm]{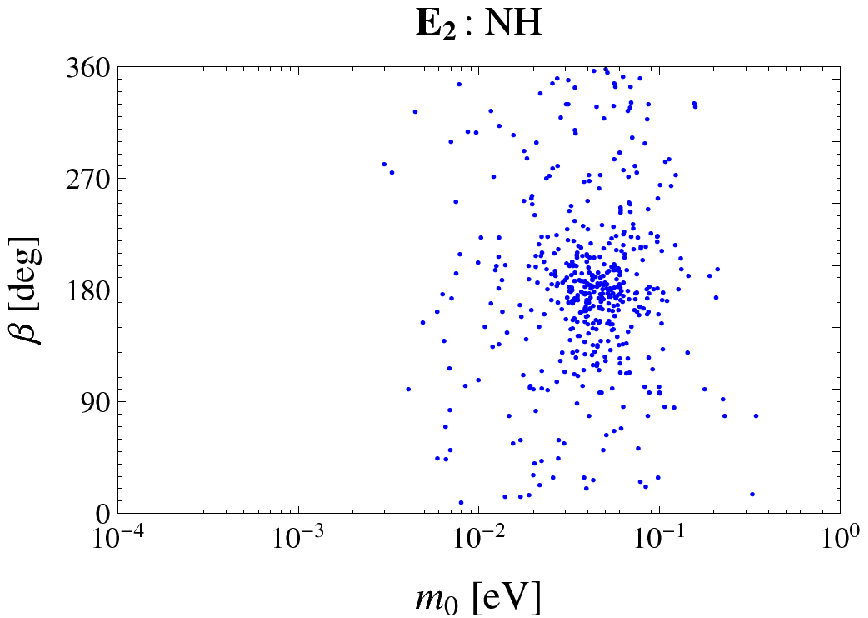} 
  \includegraphics[width=6.1cm]{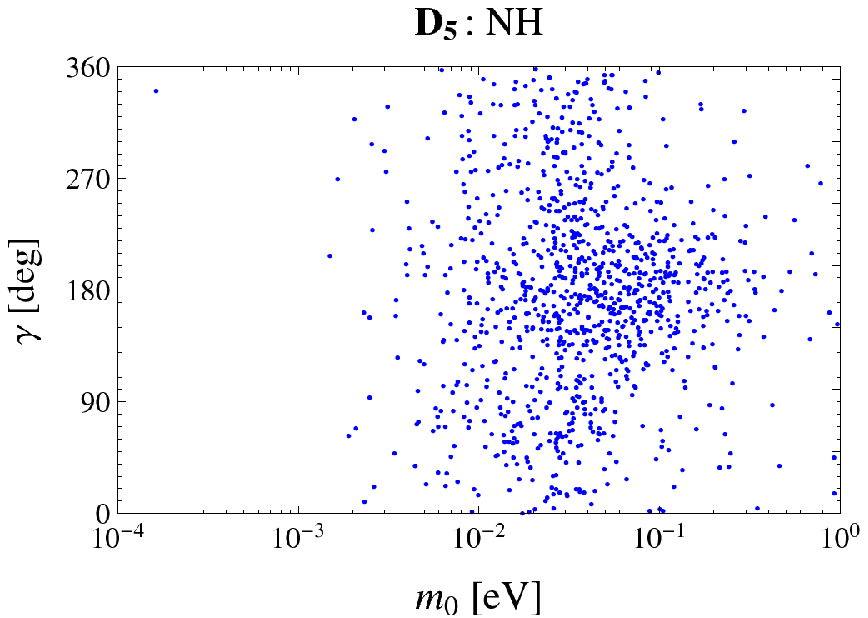} 
  \caption{Distributions of the CP violating phases $\delta_{13}$ in the texture ${\bf F}$ (NH, upper left panel), $\delta_{24}$ in ${\bf B_2}$ (NH, upper right panel), $\alpha$ in ${\bf A}$ (IH, middle left panel) and ${\bf D_5}$ (NH, middle right panel), $\beta$ in ${\bf E_2}$ (NH, lower left panel) and $\gamma$ in ${\bf D_5}$ (NH, lower right panel), as functions of the lightest neutrino mass $m_0$. The blue and red points denote, respectively, the NH and IH spectrums.}
  \label{fig:phases}
  \end{figure}

  \item
  As the analytical analysis, the numerical calculation can also predict the physical nonoscillation observables: sum of neutrino masses $\Sigma_m$, the effective neutrino mass $m_\beta$ in single beta decay and the neutrino mass $\langle m_{ee} \rangle$ in $0\nu\beta\beta$. As an explicit illustration, the three quantities for the texture ${\bf D_3}$ are presented in Fig.~\ref{fig:non-oscillation}. The three light active neutrinos are generally of order 0.01~eV and the sterile one of order 1~eV, thus the masses of the four states sum up to $\sim$1~eV, as shown in the upper panel of Fig.~\ref{fig:non-oscillation}, which is compatible with the bound on neutrino mass from cosmological observations~\cite{sn-whitepaper}.

  Numerical prediction of the effective neutrino mass $m_\beta$ in tritium beta decay for the texture ${\bf D_3}$ is presented in the lower left panel of Fig.~\ref{fig:non-oscillation}. Considering the SBL constraints, the lightest neutrino mass $m_0$ cannot too larger than 0.01~eV, or the ASMs would become very large. Thus, assuming $m_0\simeq0.01$~eV, the effective neutrino mass $m_\beta$ is around $\mathcal{O}(10^{-1}\:{\rm eV})$, consistent with the analytical prediction in Sec.~\ref{sec:analytical}. A more phenomenologically meaningful output is that the upper bound on $m_\beta$ for the NH spectrum is below the KATRIN discover sensitivity of 0.35~eV~\cite{KATRIN}, as indicated by the LO analytical expressions. If the electron neutrino mass $m_\beta$ is above 0.35~eV, KATRIN can in principle distinguish the two neutrino spectrums. However, the parameter space for the NH spectrum with $m_0\simeq0.01$~eV and $0.35\:{\rm eV} < m_\beta < 0.423\:{\rm eV}$ is rather small and it is challenging for KARTRIN to produce positive signal in this restricted region and finally pin down the neutrino spectrum. If the KATRIN experiment cannot detect electron neutrino mass but otherwise set a upper limit of 0.2~eV, then we are not able to differentiate the two neutrino hierarchies.

  For $0\nu\beta\beta$, as shown in the lower right panel of Fig.~\ref{fig:non-oscillation}, with $m_0\sim0.01$~eV, the effective Majorana neutrino mass $\langle m_{ee} \rangle$ for IH spectrum is large than $\sim$ 40 meV, consistent with the analytical predictions. The $\langle m_{ee} \rangle$ for NH spectrum is clearly much lower, predictively around 10 meV. The manifestation of nonvanishing lower limit for NH, which seems not consistent with the LO prediction, is a result from the corrections of the NLO and higher order terms. As mentioned in Sec.~\ref{sec:analytical}, its is promising for the next generation $0\nu\beta\beta$ experiments to pin down the neutrino mass spectrum and test the type II textures~\cite{0nubetabeta-review}.
  \begin{figure}[tp]
  \centering
  \includegraphics[width=6.3cm]{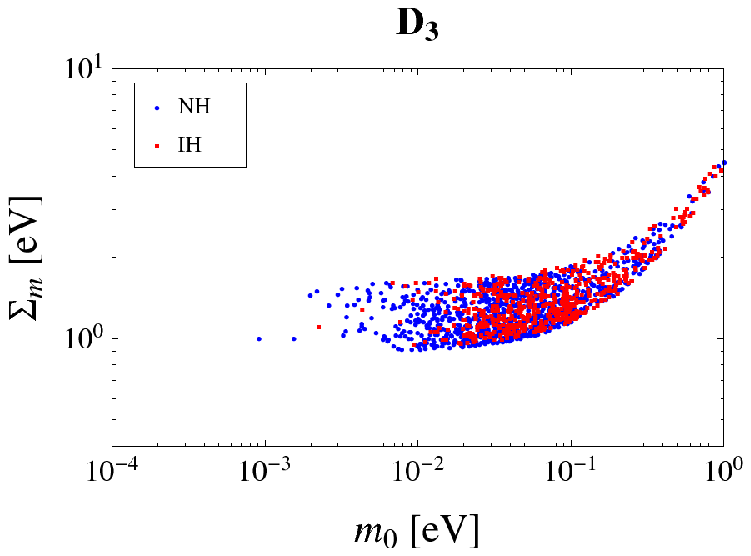} \\
  \includegraphics[width=6.3cm]{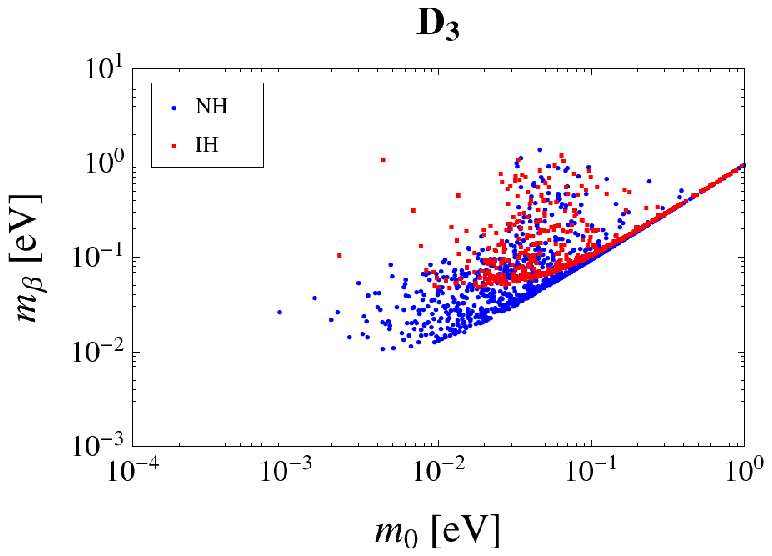} \hspace{.2cm}
  \includegraphics[width=6.3cm]{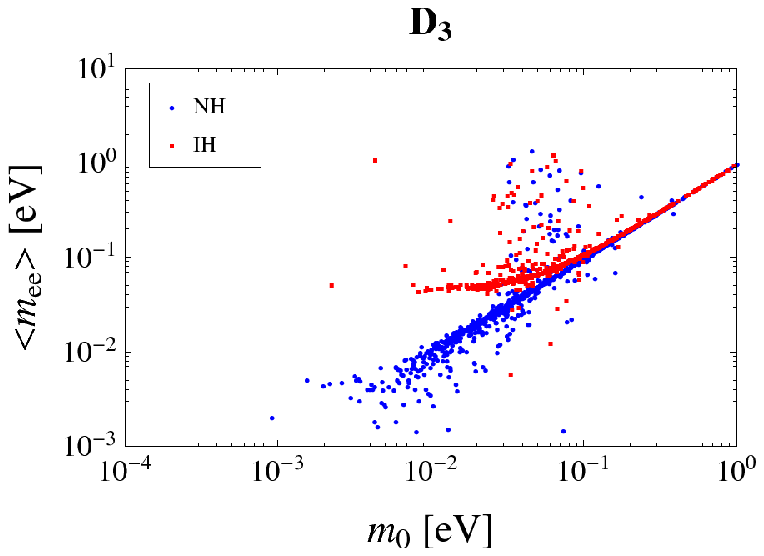} 
  \caption{Distributions of the physical nonoscillation quantities for the texture ${\bf D_3}$ as functions of the lightest neutrino mass $m_0$: sum of neutrino masses $\Sigma_m$ (upper panel), effective neutrino mass $m_\beta$ (lower left panel) in beta decay and the effective Majorana neutrino mass $\langle m_{ee} \rangle$ (lower right panel) in $0\nu\beta\beta$. The blue filled circle and red square denote, respectively, the NH and IH spectrums.}
  \label{fig:non-oscillation}
  \end{figure}
\end{itemize}
The LO analytical analysis indicates that there exist approximate $\mu-\tau$ symmetry between some of the type II textures. In the numerical calculation without any approximation, the approximate $\mu-\tau$ symmetry are reserved and the textures related by this flavor symmetry share the same phenomenological features, e.g., the preference of the textures ${\bf B_{1,\,3}}$ for the NH spectrum over IH. One antiexample is that the texture ${\bf D_5}$ prefer the CP violating phases $\delta_{24}\sim0$ and $\gamma\sim\pi$ but its partner ${\bf D_3}$ does not, as presented in Table~\ref{table:phases}. We speculate this explicit and significant flavor ``asymmetry'' is due to the NLO and higher order corrections beyond Eq.~(\ref{eqn:II}).

\section{Textures without CP violation}
\label{sec:cpconservation}

When all the CP violating phases vanish, i.e., $\delta_{13,\,14,\,24}=\alpha=\beta=\gamma=0$, without tuning of the phases it would be more challenging for the vanishing conditions (\ref{3zero}) to be satisfied. For instance, the CP conserving partners of the six vanishing conditions in the active sector of $M_\nu$ are, respectively,
\begin{subequations}
\begin{align}
m_{ee} &: \;  m_4 s_{14}^2
    +\Big( m_1 c_{12}^2
    +m_2 s_{12}^2 \Big) =0 \,, \label{eqn:eecpc}\\
m_{\mu\mu} &:\;  m_4 s_{24}^2
    + \Big( m_1 s_{12}^2 c_{23}^2
    +m_2 c_{12}^2 c_{23}^2
    +m_3 s_{23}^2  \Big) =0 \,, \label{eqn:mumucpc}\\
m_{\tau\tau} &:\;   m_4 s_{34}^2
   + \Big( m_1 s_{12}^2 s_{23}^2
   +m_2 c_{12}^2 s_{23}^2
   +m_3 c_{23}^2  \Big) =0 \,, \label{eqn:tautaucpc}\\
m_{e\mu} &:\;  m_4 s_{14} s_{24}
   - \Big( (m_1-m_2) s_{12}c_{12}c_{23} -m_3 s_{23}s_{13}  \Big) =0 \,, \label{eqn:emucpc} \\
m_{e\tau} &:\;  m_4 s_{14} s_{34}
   + \Big( (m_1-m_2) s_{12}c_{12}s_{23}
   +m_3 c_{23} s_{13}  \Big)  =0 \,, \label{eqn:etaucpc} \\
m_{\mu\tau} &:\;  m_4 s_{24} s_{34}
   -s_{23}c_{23} \Big( m_1 s_{12}^2
    +m_2 c_{12}^2
    -m_3  \Big) =0 \,. \label{eqn:mutaucpc}
\end{align}
\end{subequations}
Obviously, the first four equations can never be satisfied, as the left-hand sides of them are definitely positive. Consequently, none of the 20 type II textures without CP violation can even be compatible with the current data on active neutrino oscillations. We calculate numerically all these 20 CP conserving textures and  predictively obtain the negative result. For the 100 type I CP conserving textures, due to the same reason above for the type II textures, they are all expected to be excluded by the active neutrino data. We also numerically analyze these 100 CP conserving textures and again arrive at the same conclusion.

\section{Conclusion}
\label{sec:conclusion}

The LSND anomaly, MiniBooNE experiment, the reactor anomaly, the Gallium anomaly and cosmological observations all point to the existence of light eV scale sterile neutrino(s). Texture zero is a powerful framework to investigate the phenomenologies of the neutrino sector of the SM. In this work we apply the framework of texture zeros to the smallest extension of the neutrino sector of SM, i.e., the (3+1) model, and examine all the 120 textures of $4\times4$ Majorana neutrino mass matrix $M_\nu$ with three zeros. Although all the $3\times3$ neutrino matrices with three zeros have been excluded, we find that when the neutrino sector is extended to include one sterile flavor, the situation is completely different.

We classify the 120 textures into two large classes, the 100 type I textures with zeros in the sterile sector (the fourth row and column) of $M_\nu$ and the 20 type II textures without zero in the sterile sector. With the approximations of large mass hierarchy $m_{1,\,2,\,3} \ll m_4$ and smallness of mixings $\theta_{13,\,14,\,24,\,34} \ll 1$, all the analytical expressions can be greatly simplified. We analyze both analytically and numerically all the 120 textures and find that the 100 type I textures are all excluded by current neutrino data, mainly due to the large active-sterile mass splitting. The exception among the type I textures are the class $\mathcal{E}$ textures with only $m_{\tau s}=0$ in the sterile sector, but they are highly disfavored by experiments as a result of the extremely small available parameter space. The 20 type II textures with all the three zeros in the active sector are phenomenologically much more favored. As for the textures within the three-neutrino scheme~\cite{zero-permutation}, some of the textures are related by the approximate $\mu-\tau$ symmetry~\cite{2zero-sterile} and the symmetric partners always share the same phenomenological features. Their remarkable features revealed by both the analytical and numerical analysis are collected as follows:
\begin{enumerate}
  \item
  To obtain vanishing elements in the active sector of $M_\nu$ and keep compatible with the $\mathcal{O}(0.1)$ active-sterile mixing, the active neutrino mass and lightest neutrino mass are required to be of order 0.01~eV.
  \item
  We can also state the same fact in another way: If the active neutrinos are of order 0.01~eV, we naturally arrive at the ASMs are of order 0.1. That is to say, the 20 textures are overall compatible with current experimental data, which is verified by both analytical and numerical analysis. Numerical calculations reveal further that among the 20 textures, only the texture ${\bf C_3}$ is actually excluded and the other 19 are more or less compatible with data.
  \item
  Both the two neutrino spectrums of NH and IH for the three light active flavors are allowed in the textures ${\bf A}$, ${\bf B_{2,\,4,\,5,\,6}}$, ${\bf D_{3,\,4,\,5,\,6}}$, ${\bf E_{2,\,3}}$, ${\bf F}$, whereas the seven textures ${\bf B_{1,\,3}}$, ${\bf C_{1,\,2}}$, ${\bf D_{1,\,2}}$, ${\bf E_{1}}$ prefer much for the NH spectrum over IH.
  \item
  Some of the CP violating phases in the 19 textures exhibit preference of particular values, e.g., the Dirac phase $\delta_{13}$ in the texture ${\bf F}$ favors the values around $\pi/2$ or $3\pi/2$, and in ${\bf A}$ the Majorana phase $\alpha\sim\pi$. All the predictions on the phases are collected in Table~\ref{table:phases}. The Dirac-type phases, especially the phase $\delta_{13}$, can be observed in the upcoming long baseline neutrino oscillation experiments, while the Majorana phases are crucial to the $0\nu\beta\beta$ experiments. All these relevant oscillation and nonoscillation experiments are promising to examine the variant phenomenological implications and test these textures in the (3+1) model.
  \item
  The 0.01 eV active neutrinos and eV scale sterile neutrino sum up to $\sim$eV, compatible with the cosmological bound on neutrino mass. No matter whether the KATRIN beta decay experiment can observe the electron neutrino mass or not, it is challenging for it to distinguish the two neutrino spectrums in the three-zero textures. In contrast, it is promising for the next generation $0\nu\beta\beta$ experiments with sensitivity of order 10 meV can clearly differentiate NH from IH and then test these textures. For example, if the effective Majorana neutrino mass comes up to be 30 meV, the three active neutrino are probably of inverted hierarchy and then those textures preferable of NH would be almost excluded.\footnote{The $0\nu\beta\beta$ test of the textures in this item is based on the effective neutrino mass, while the $0\nu\beta\beta$ test in the previous item is related to the Majorana phase detection in $0\nu\beta\beta$ experiments. Although both the tests are in principle feasible, the latter one is much more challenging than the former~\cite{0nubetabeta-review}.}
\end{enumerate}
In addition to the textures with the most general CP violation, we calculate also the textures without any CP violation and find that without the tuning of the phases in the correlation equations, none of the 120 textures are allowed by current neutrino data.

In the end we stress that with more accurate neutrino data, especially the determination of the octant of $\theta_{23}$~\cite{octant}, the textures with three zeros in the (3+1) model will be more severely constrained. It is likely that more textures will be excluded and we can obtain more accurate phenomenological predictions, e.g., the effective neutrino mass in the neutrinoless double beta decay.

\begin{acknowledgments}
The author would like to thank Wei Chao for the frequent discussions in the early stage of this work. The author also appreciates Zhi-zhong Xing, Rabindra Mohapatra, and Guo-Zhu Ning for reading the manuscript and the valuable comments by Xing and Ning. This work is supported by National Natural Science Foundation of China (NSFC) under Grant No. 11105004.
\end{acknowledgments}

\appendix
\section{Classification of the 120 textures}
\label{sec:classification}

The number of textures for $4\times4$ $M_\nu$ with three zeros is rather large; thus, it is beneficial for us to properly classify and label the 120 distinct textures. As the sterile neutrino is essentially different from the three active ones,
the 120 textures can be roughly classified into two large categories: (I) 100 textures with zeros in the sterile sector of $M_\nu$ and (II) 20 textures without vanishing elements in the sterile sector. As we will see in the analytical and numerical discussions, predictions of the structures in the two classes on the ASM angles are completely different.

For the convenience of discussions, the 100 textures of $M_\nu$ of class I can be further classified into six subgroups $\mathcal{A}-\mathcal{F}$, according to the number of zeros in the sterile sector. The matrices given below are the sample matrices for the six subgroups:
\begin{eqnarray}
\label{matrix:I}
  && \mathcal{A}:\;
  \left(\begin{array}{cccc}
  \times & \times & \times & 0 \\ \times& \times & \times & 0 \\  \times& \times & \times & \times \\  {0} & 0 & \times & 0
  \end{array} \right) \,;\qquad
  \mathcal{B}:\;
  \left(\begin{array}{cccc}
      \times & \times & \times & {0} \\  \times & \times & \times & {0} \\  \times & \times & \times & {0} \\  {0} & {0} & {0} & {\times}
  \end{array} \right) \,;\qquad
  \mathcal{C}:\;
  \left(\begin{array}{cccc}
      0 & \times & \times & 0 \\  \times & \times & \times & \times \\ \times& \times& \times & \times \\   0&\times&\times& 0
  \end{array} \right) \,;\nonumber \\
&&  \mathcal{D}:\;
  \left(\begin{array}{cccc}
      0 & \times & \times & 0 \\  \times& \times & \times & 0 \\ \times&\times& \times & \times \\  0&0&\times& \times
  \end{array} \right) \,;\qquad
  \mathcal{E}:\;
  \left(\begin{array}{cccc}
      0 & 0 & \times & \times \\  0& \times & \times & \times \\  \times&\times& \times & \times \\  \times&\times&\times& 0
  \end{array} \right) \,;\qquad
  \mathcal{F}:\;
  \left(\begin{array}{cccc}
      0 & 0& \times & 0 \\  0& \times & \times & \times \\  \times&\times& \times & \times \\  0&\times&\times& \times
  \end{array} \right) \,.
\end{eqnarray}
All the three texture zeros of the first two groups $\mathcal{A}$ and $\mathcal{B}$ are in the sterile sector. The three textures of type $\mathcal{A}$ have the fourth diagonal element $m_{ss}=0$, whereas all the three ASM angles of the texture of type $\mathcal{B}$ are zeros and this texture is apparently excluded by the LSND experiment. The two following groups $\mathcal{C}$ and $\mathcal{D}$ consist of the $M_\nu$ structures with one zero in the active sector and the other two in the sterile sector. $m_{ss}=0$ holds true for the textures of type $\mathcal{C}$, while for the textures of type $\mathcal{D}$, the two texture zeros in the sterile sector are both nondiagonal elements. For the last two groups $\mathcal{E}$ and $\mathcal{F}$, only one of the three texture zeros are from the sterile sector. For type $\mathcal{E}$ $m_{ss}=0$, whereas for type $\mathcal{F}$ one of the three nondiagonal elements in the fourth row or column vanishes.

The 20 patterns of $M_\nu$ with all the three zeros in the active sector can also be further classified into subgroups. $3\times3$ neutrino matrices with three texture zeros have been classified according to the positions of zeros, or more specifically according to the number of vanishing diagonal entries~\cite{Xing}. The 20 $4\times4$ matrices are analogically classified into six subgroups and they are explicitly labeled as follows:
\begin{eqnarray}
  && {\bf A}:\;
  \left( \begin{array}{cccc} 0 & \times & \times & \times \\ \times  & 0 & \times & \times \\  \times & \times & 0 & \times \\ \times  & \times & \times & \times  \end{array} \right) \,;
\end{eqnarray}
\begin{eqnarray}
  && {\bf B_1}:\;
  \left( \begin{array}{cccc} 0 & \times & 0 & \times \\ \times  & 0 & \times & \times \\ 0 & \times & \times & \times \\  \times & \times & \times & \times  \end{array} \right) ,\quad
  {\bf B_2}:\;
  \left( \begin{array}{cccc} 0 & \times & \times & \times \\ \times & 0 & 0 & \times \\ \times & 0 & \times & \times \\  \times & \times & \times & \times \end{array} \right) ,\quad
  {\bf B_3}:\;
  \left( \begin{array}{cccc} 0 & 0 & \times & \times \\ 0  & \times & \times & \times \\ \times & \times & 0 & \times \\  \times & \times & \times & \times  \end{array} \right) ,\quad \nonumber \\
&&  {\bf B_4}:\;
  \left( \begin{array}{cccc} 0 & \times & \times & \times \\  \times & \times & 0 & \times \\ \times & 0 & 0 & \times \\ \times & \times & \times & \times  \end{array} \right) ,\quad
  {\bf B_5}:\;
  \left( \begin{array}{cccc} \times & 0 & \times & \times \\ 0  & 0 & \times & \times \\ \times &\times& 0 & \times \\  \times&\times&\times& \times  \end{array} \right) ,\quad
  {\bf B_6}:\;
  \left( \begin{array}{cccc} \times & \times & 0 & \times \\  \times & 0 & \times & \times \\ 0 &\times& 0 & \times \\ \times &\times&\times& \times  \end{array} \right) \,;
\end{eqnarray}
\begin{eqnarray}
  && {\bf C_1}:\;
  \left( \begin{array}{cccc} 0 & 0 & \times & \times \\ 0  & 0 & \times & \times \\ \times &\times& \times & \times \\ \times &\times&\times& \times  \end{array} \right) ,\quad
  {\bf C_2}:\;
  \left( \begin{array}{cccc} 0 & \times & 0 & \times \\ \times & \times & \times & \times \\ 0 &\times& 0 & \times \\  \times &\times&\times& \times \end{array} \right) ,\quad
  {\bf C_3}:\;
  \left( \begin{array}{cccc} \times & \times & \times & \times \\  \times & 0 & 0 & \times \\ \times &0& 0 & \times \\  \times&\times&\times& \times  \end{array} \right) \,;
\end{eqnarray}
\begin{eqnarray}
  && {\bf D_1}:\;
  \left( \begin{array}{cccc} 0 & 0 & \times & \times \\ 0  & \times & 0 & \times \\ \times &0& \times & \times \\  \times&\times&\times& \times  \end{array} \right) ,\quad
  {\bf D_2}:\;
  \left( \begin{array}{cccc} 0 & \times & 0 & \times \\ \times & \times & 0 & \times \\ 0 &0& \times & \times \\   \times&\times&\times& \times \end{array} \right) ,\quad
  {\bf D_3}:\;
  \left( \begin{array}{cccc} \times & 0 & 0 & \times \\ 0  & 0 & \times & \times \\ 0 &\times& \times & \times \\  \times&\times&\times& \times  \end{array} \right) ,\quad \nonumber \\
&&  {\bf D_4}:\;
  \left( \begin{array}{cccc} \times & \times & 0 & \times \\ \times  & 0 & 0 & \times \\ 0 &0& \times & \times \\  \times&\times&\times& \times  \end{array} \right) ,\quad
  {\bf D_5}:\;
  \left( \begin{array}{cccc} \times & 0 & 0 & \times \\  0 & \times & \times & \times \\ 0 &\times& 0 & \times \\  \times&\times&\times& \times  \end{array} \right) ,\quad
  {\bf D_6}:\;
  \left( \begin{array}{cccc} \times & 0 & \times & \times \\  0 & \times & 0 & \times \\ \times &0& 0 & \times \\ \times &\times&\times& \times  \end{array} \right) \,;
\end{eqnarray}
\begin{eqnarray}
  && {\bf E_1}:\;
  \left( \begin{array}{cccc} 0 & 0 & 0 & \times \\  0 & \times & \times & \times \\ 0 &\times& \times & \times \\ \times &\times&\times& \times  \end{array} \right) ,\quad
  {\bf E_2}:\;
  \left( \begin{array}{cccc} \times & 0 & \times & \times \\ 0 & 0 & 0 & \times \\ \times &0& \times & \times \\  \times &\times&\times& \times \end{array} \right) ,\quad
  {\bf E_3}:\;
  \left( \begin{array}{cccc} \times & \times & 0 & \times \\  \times & \times & 0 & \times \\ 0 &0& 0 & \times \\  \times&\times&\times& \times  \end{array} \right) \,;
\end{eqnarray}
\begin{eqnarray}
  && {\bf F}:\;
  \left( \begin{array}{cccc} \times & 0 & 0 & \times \\ 0  & \times & 0 & \times \\ 0 & 0 & \times & \times \\  \times&\times&\times& \times  \end{array} \right) \,.
\end{eqnarray}
All the first three diagonal elements of the texture {\bf A} vanish, whereas two of the first three diagonal elements of the textures of types {\bf B} and {\bf C} are zeros, only one for types {\bf D} and {\bf E} are vanishing and none of the three diagonal elements in the active part of {\bf F} are zero. The three textures of type {\bf C} are of rank three, which are distinguishable from the six of type {\bf B} which are of rank four. Similarly, the six structures of type {\bf D} are of rank four, whereas the three of type {\bf E} are of rank three.



\end{document}